\documentclass[5p,twocolumn]{elsarticle}

\usepackage[switch,columnwise]{lineno}
\modulolinenumbers[200]

\journal{Information and Software Technology}

\biboptions{sort&compress}

\usepackage{makecell}
\usepackage{booktabs} 
\usepackage[utf8]{inputenc} 
\usepackage[dvipsnames]{xcolor}
\usepackage[inline]{enumitem}
\usepackage{amsmath}
\usepackage{subcaption}
\usepackage{multirow}
\usepackage{multicol}
\usepackage{microtype}
\usepackage{alltt}
\usepackage{listings}
\usepackage{tabu}

\usepackage{amsthm}

\usepackage{framed}
\usepackage{xspace}
\PassOptionsToPackage{hyphens}{url}\usepackage{hyperref}  
\usepackage
{cleveref}  

\usepackage{letltxmacro}

\usepackage[disable]{todonotes}

\LetLtxMacro{\oldtodo}{\todo}
\renewcommand{\todo}[1]{\oldtodo[inline]{#1}}

\usepackage{pgfplotstable}
\usepackage{pgfplots}
\definecolor{lightgreen}{rgb}{0.6,0.92,0.5}
\definecolor{green}{rgb}{0.10,0.80,0}
\definecolor{darkgreen}{rgb}{0,0.37,0}
\definecolor{lightblue}{rgb}{0.5,0.85,0.90}
\definecolor{blue}{rgb}{0,0.4,0.80}
\definecolor{darkblue}{rgb}{0,0.2,0.40}
\newcommand*{\plotscale}{0.90} 

\newcommand{\VUDENC}{\textnormal{\textsc{Vudenc}}\xspace}

\begin{document}
\makeatletter
\def\ps@pprintTitle{%
	\let\@oddhead\@empty
	\let\@evenhead\@empty
	\def\@oddfoot{\parbox[t]{1\linewidth}{\footnotesize \textit{Accepted Manuscript. \@journal \\ December 09, 2021 \\ ~\url{https://doi.org/10.1016/j.infsof.2021.106809} \\
	License: \href{https://creativecommons.org/licenses/by-nc-nd/4.0/}{CC BY-NC-ND 4.0} }}}%
	\let\@evenfoot\@oddfoot}
\makeatother

\begin{frontmatter}

\title{VUDENC: Vulnerability Detection with Deep Learning\\on a Natural Codebase for Python}

\author[HUB]{Laura Wartschinski}
\author[HUB]{Yannic Noller}
\author[HUB,UPB]{Thomas Vogel}
\author[HUB,UB]{Timo Kehrer}
\author[HUB]{Lars Grunske}

\address[HUB]{Department of Computer Science, Humboldt-Universität zu Berlin, Unter den Linden 6, 10099 Berlin, Germany\\
	\{wartschinski,noller,thomas.vogel,kehrer,grunske\}@informatik.hu.berlin.de}
\address[UPB]{Department of Computer Science, University of Paderborn, Warburger Straße 100, 33098 Paderborn, Germany}
\address[UB]{Department of Computer Science, University of Bern, Hochschulstrasse 6, 3012 Bern, Switzerland}

\begin{abstract}
{\bf Context:}
Identifying potential vulnerable code is important to improve the security of 
our software systems. 
 However, the manual detection of software vulnerabilities requires expert 
knowledge and is time-consuming, and must be supported by automated techniques.

\vspace{0.1cm}\noindent{\bf Objective:}
Such automated vulnerability detection techniques should achieve a high 
accuracy, point developers directly to the vulnerable code fragments, scale to 
real-world software, generalize across the boundaries of a specific software 
project, and require no or only moderate setup or configuration effort.

\vspace{0.1cm}\noindent{\bf Method:} 
In this article, we present \VUDENC (Vulnerability Detection with Deep Learning 
on a Natural Codebase), a deep learning-based vulnerability detection tool that 
automatically learns features of vulnerable code from a large and real-world 
Python codebase. 
 \VUDENC applies a word2vec model to identify semantically similar code tokens 
and to provide a vector representation. 
 A network of long-short-term memory cells (LSTM) is then used to classify 
vulnerable code token sequences at a fine-grained level, highlight the specific 
areas in the source code that are likely to contain vulnerabilities, and provide 
confidence levels for its predictions.

\vspace{0.1cm}\noindent{\bf Results:} 
To evaluate \VUDENC, we used 1,009 vulnerability-fixing commits from different
GitHub repositories that contain seven different types of vulnerabilities 
(SQL injection, XSS, Command injection, XSRF, Remote code execution, Path 
disclosure, Open redirect) for training. 
 In the experimental evaluation, \VUDENC achieves
a recall of 78\%-87\%, a precision of 82\%-96\%, and an F1 score of 80\%-90\%. 
 \VUDENC's code, the datasets for the vulnerabilities, and the Python 
corpus for the word2vec model are available for reproduction.

\vspace{0.1cm}\noindent{\bf Conclusions:}
Our experimental results suggest that \VUDENC is capable of outperforming most 
of its competitors in terms of vulnerably detection capabilities on real-world
software. 
 Comparable accuracy was only achieved on synthetic benchmarks, within single 
projects, or on a much coarser level of granularity such as entire source code 
files.
\end{abstract}

\begin{keyword}
Static Analysis, Vulnerability Detection, Deep Learning, Long-Short-Term Memory Network, Natural Codebase, Software Repository Mining
\end{keyword}

\end{frontmatter}

\linenumbers

\section{Introduction}
\label{sec:intro}

Software vulnerabilities can make a software-intensive system an easy target for cyber-attacks whose consequences may be severe, ranging from information loss or disclosure of secret information to manipulation and system failure. All of them may have a serious impact on businesses, governments, society and individuals. An exploit like the ransomware \textit{WannaCry}, for example, led to the shutdown of hospitals, telecommunication services and transportation systems, and caused damage in the order of hundreds of millions of dollars~\cite{DanGoodin.2017}.

Many vulnerabilities are caused by subtle code flaws, spanning a few or even just a single line of code~\cite{Yamaguchi.2012}. For instance, the famous \textit{Heartbleed} bug, a vulnerability in the OpenSSL cryptographic library that affected billions of internet users, could have been prevented with two more lines of code~\cite{Durumeric.2014}. To mitigate such flaws, constructive approaches to secure software engineering (e.g., model-based secure software design~\cite{jurjens2002umlsec,peldszus2021ontology,burger2020ontology} or up-to-date collections of well-known vulnerabilities and security guidelines~\cite{CVE}), need to be complemented by analytical quality assurance techniques for detecting vulnerabilities in source code.

Given the increasing size and complexity of contemporary software as well as the growing number of potential attacks, the manual detection of vulnerabilities in source code is almost impossible, and must be supported by \mbox{(semi-)}automated techniques~\cite{arkin2005software,Zimmermann.2010,Shin.2010,Shin.2013,kim2017vuddy,li2016vulpecker,Scandariato.2014,Morrison.2015,Russell.2018,Pang.2015,Ma.2017,Li.2018,Dam.2017,Hovsepyan.2012}. Such techniques should (i)~achieve a high accuracy in detecting potential vulnerabilities, and (ii)~point developers directly to the vulnerable code fragments. At the same time, the techniques should be (iii)~scalable to real-world software, (iv)~generalize across the boundaries of a specific software project, and (v)~require no or only moderate setup or configuration effort. However, none of the existing techniques fulfills all of these requirements.

Dynamic program analysis approaches, notably software penetration testing~\cite{arkin2005software}, require a wide range of representative test cases defined by security experts who need to know attacking strategies and reason about the software under test like an attacker~\cite{Dam.2017,Pang.2015}. Traditional static analysis tools are rule-based systems which rely on the definition of features of vulnerable code. On the one hand, defining such features with the help of human experts is a tedious task which is prone to errors yielding incomplete rule sets~\cite{Li.2018,Russell.2018}. On the other hand, the use of generic features such as software metrics~\cite{Zimmermann.2010,Shin.2010,Shin.2013} suffers from high false positive rates~\cite{austin2011one,ceccato2016static,Ghaffarian.2017}, while experiments with structural approaches based on code clone detection~\cite{kim2017vuddy} and similarity search~\cite{li2016vulpecker} expose false negative rates which are too high to be considered acceptable~\cite{Li.2018}.

With the increasing availability of open-source repositories, it has been suggested to use a data-driven approach to vulnerability detection. Various machine learning (ML) techniques have been applied to learn vulnerable features of code, however, with varying success. Simple bag-of-words and classification algorithms~\cite{Scandariato.2014,Morrison.2015} lead to rather disappointing accuracy since they are not able to capture the sequential nature and semantic structure of source code~\cite{Dam.2017}. Experimental results obtained for more sophisticated approaches~\cite{Russell.2018,Pang.2015,Ma.2017,Li.2018,Dam.2017,Hovsepyan.2012}, including different kinds of deep neural networks (DNNs), achieve accuracy values of more than 80\% in terms of precision and recall. However, many studies work on synthetic code examples~\cite{Russell.2018,Li.2018} or are only applicable to a small set of projects~\cite{Pang.2015,Dam.2017,Hovsepyan.2012}. Furthermore, many proposed approaches only classify whole files~\cite{Pang.2015,Dam.2017,Hovsepyan.2012} or API calls~\cite{Li.2018}, which is easier to achieve but too coarse-grained to point developers directly to vulnerable code fragments. Finally, the labeling of training datasets is often not fully automated, but relies on manual intervention~\cite{Ma.2017,Li.2018}, which does not scale to large training datasets.
In summary, existing solutions are far from reaching an optimal trade-off between accuracy, guidance, applicability, generalizability, and configuration effort, as expressed by the above requirements (i) to (v).
\textit{The challenge of how to overcome this limitation is the overall research objective of our work}.

In this paper, we present \VUDENC (Vulnerability Detection with Deep Learning on a Natural Codebase), a deep learning-based vulnerability detection system that automatically learns features of vulnerable code from a large and real-world codebase. To ensure its applicability to a wide range of different kinds of software, \VUDENC works on a textual representation of source code. Moreover, our choice of a textual representation is motivated by recent outstanding research results in the field of natural language processing (NLP) using long-short-term-memory networks (LSTMs)~\cite{Hochreiter.1997}, relying on the assumption that programming languages and natural languages are fundamentally similar in nature. As it is customary in NLP, source code is treated as a sequence of tokens (identifiers, keywords, literals, operators, etc.), which are then embedded in a numerical vector space amenable to an LSTM. Embeddings are obtained using word2vec~\cite{Word2Vec}, which has been successfully applied to similar problems before~\cite{Liu.2018}. Tokens are grouped into code fragments of configurable size to allow for the context of a token to be taken into account, while the classification itself is very fine-granular, pointing developers to the specific code fragment that is potentially vulnerable. Labeled training datasets are obtained in a fully automated fashion by crawling for security-related fixes in the commit history of a software repository. We  implemented our approach for Python which, despite its steadily increasing popularity and widespread usage in many application fields, has not been the focus of any research on vulnerability detection with deep learning techniques. Experimental subjects were extracted from GitHub, which hosts a huge number of real-world Python projects.

In summary, this paper makes the following contributions:
\begin{itemize}
\item A vulnerability detection approach based on deep learning which is generally applicable on textual representations of source code, pointing developers directly to potentially vulnerable code fragments.
\item A fully automated labeling approach that can be applied to a very large corpus of real-world software without the need for manual intervention.
\item An implementation of the overall approach in a prototypical tool called \VUDENC which has been trained on a labeled dataset comprising security fixes of 14,686 Python projects hosted on GitHub.
\item Experimental results that achieved a recall of 78\%-87\% and precision of 82-96\%, suggesting that \VUDENC can outperforming its direct competitors. Comparable accuracy was only achieved on synthetic benchmarks, within single projects, or on a much coarser level of granularity.
\end{itemize}

\section{Related Work}
\label{sec:relwork}
\begin{figure}[t]
	\centering
	\includegraphics[width=\linewidth]{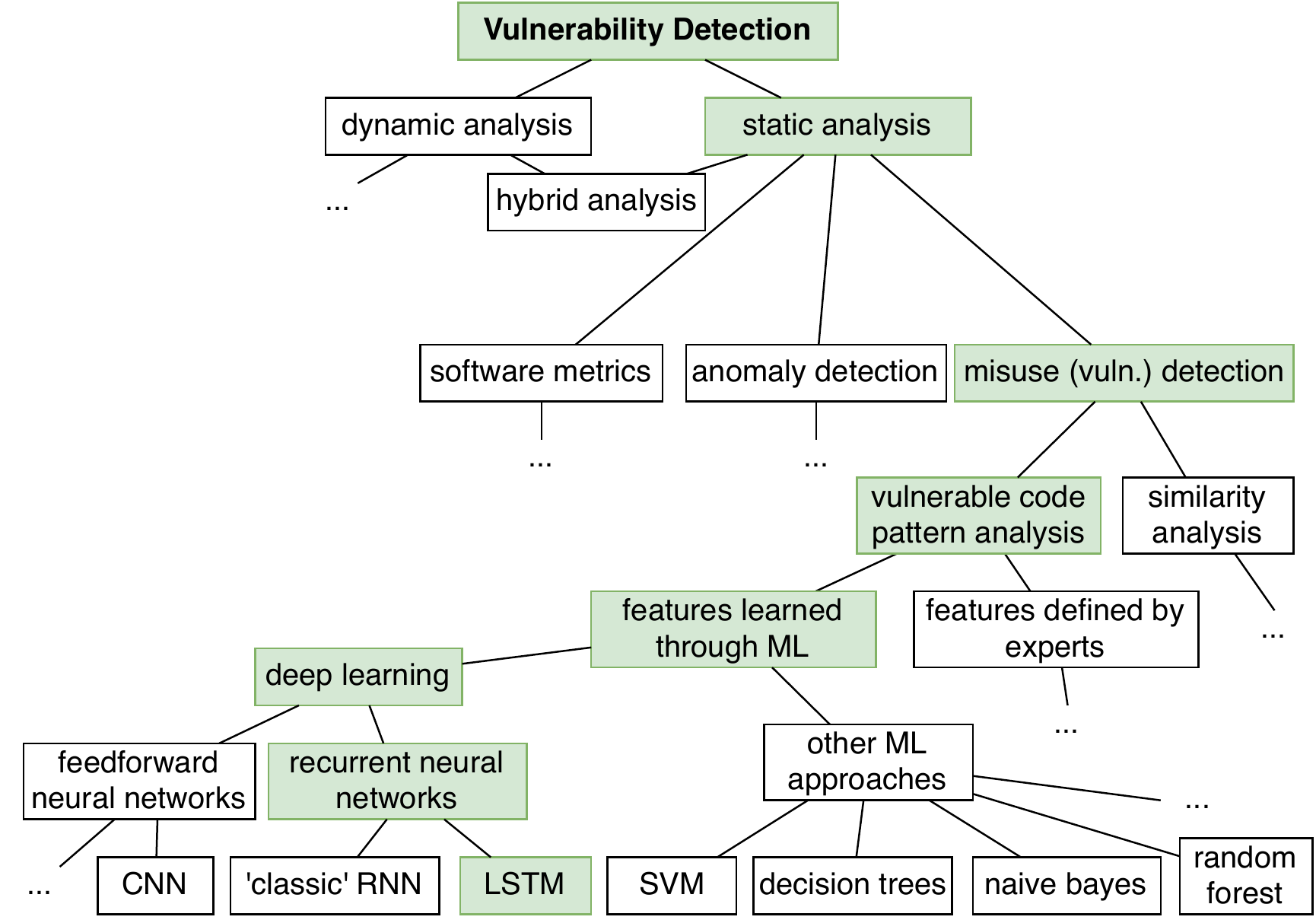}
	\caption{Classification of approaches for vulnerability detection.}
	\label{fig:overview}
\end{figure}

We discuss related work on vulnerability detection, especially approaches that use static analysis for this purpose, similarly to \VUDENC. As depicted in Figure~\ref{fig:overview}, such approaches can be classified into three groups, depending on whether they are based on software metrics, anomaly detection, or misuse detection. We discuss these three groups and position \VUDENC in this context (cf.\ green elements in Figure~\ref{fig:overview}).

\subsection{Vulnerability detection based on software metrics}

To detect or predict vulnerabilities, the most commonly used features are software and developer metrics such as lines of code (LOC), cyclomatic complexity, code churn, developer activity, coupling, number of dependencies or legacy metrics~\cite{Morrison.2015}.
Such metrics have been used as features for building fault prediction models~\cite{Hall.2011,Nagappan+2006}. For instance, Nagappan et al.~\cite{Nagappan.2008} use organizational metrics to predict faults in software.
However, such metrics do not generalize well across projects~\cite{Nagappan+2006} and do not capture the semantics of the code~\cite{Shin.2008}.
In contrast, they rather apply a foregone conclusion that certain meta-features will be related to vulnerabilities, which is not necessarily true~\cite{Hovsepyan.2012}.

Nonetheless, code metrics have been used with machine learning (ML) to predict software vulnerabilities.
Shin et al.~\cite{Shin.2008} use nine complexity metrics to predict vulnerabilities in Javascript projects, achieving a low false positive rate, but a relatively high false negative rate.
Later, the authors leverage code complexity, code churn and developer metrics to predict vulnerabilities with linear discriminant analysis and Bayesian networks~\cite{Shin.2010}.
Based on complexity, coupling and cohesion metrics (CCC), Chowdhury et al.~\cite{Chowdhury.2011} predict vulnerabilities with fault detection techniques.
They study releases of Mozilla Firefox and use decision trees, random forest, logistic regression, and naive Bayes models to predict vulnerabilities.
Zimmerman et al.~\cite{Zimmermann.2010} consider more metrics (code churn, code complexity, code coverage, organizational measures and actual dependencies) and found a weak yet statistically significant correlation between them while focusing on the proprietary code of Windows Vista. They use logistic regression to predict vulnerabilities based on these metrics.
Neuhaus et al.~\cite{Neuhaus.2007} look at import statements in the Mozilla project, using support vector machines for the sake of prediction.
Finally, other researchers made predictions based on commit messages.
Zhou et al.~\cite{Zhou.2017} leverage a K-fold stacking algorithm to analyze commit messages to successfully predict whether a commit contains vulnerabilities. In contrast, Russel et al.~\cite{Russell.2018} found that both humans and machine learning algorithms performed poorly at predicting build failures or bugs from commit messages.

In contrast to all of these approaches, \VUDENC does not take meta-information in the form of software metrics into account, but learns features directly from the source code.

\subsection{Vulnerability detection based on anomaly detection}

Anomaly detection refers to the problem of identifying deviations from normal or expected patterns in source code. It assumes that code fragments not conforming to those patterns can cause a defect. To extract such code patterns, data-mining techniques have been used on source code.
For instance, Li et al.~\cite{Li.2005} developed PR-Miner to find code patterns in any programming language by associating code fragments that are frequently used together. Violations reported by PR-Miner have been confirmed as bugs in Linux, PostgreSQL and the Apache HTTP Server.
However, a fundamental problem is that bugs that are themselves typical patterns, and therefore occur frequently, are systematically overlooked~\cite{Yamaguchi.2012}. At the same time, rare programming patterns or API usages can be flagged as false positive as they do not occur often.
Consequently, anomaly detection approaches often have high false-positive rates~\cite{Ghaffarian.2017}.

\VUDENC differs from anomaly detection as it uses explicit labels on code fragments to train a model on vulnerable and invulnerable code instead of identifying ``typical'' code fragments. In other words, we do not assume that ``typical'' code fragments are also invulnerable fragments.

\subsection{Vulnerability detection based on misuse detection}

Vulnerable code pattern analysis and similarity analysis try to identify typical characteristics of vulnerabilities in order to detect vulnerable code.
In vulnerable code pattern analysis, vulnerable code segments are analyzed with ML techniques to extract their typical features. Those features represent patterns that can be applied on new code segments to find vulnerabilities. Most of the approaches in this area gather a large dataset, process this data to extract feature vectors, and run ML on this data~\cite{Ghaffarian.2017}.
In similarity analysis, a vulnerable code snippet is given and the goal is to find similar ones, assuming that they are at risk to share the vulnerability. This kind of analysis is well-suited for nearly identical code in which the inherent structure of the compared code fragments is very similar~\cite{Li.2018}.
In both kinds of analysis, features of the code are obtained partially or fully automatically, eliminating the need for subjective human experts. By learning directly from a dataset of code which includes both vulnerable and non-vulnerable code fragments, an unbiased model can be built.

Approaches in this area differ from each other in many aspects:
\textit{Dataset} (synthetic or real-life data), 
\textit{size of the dataset} (e.g., number of projects, classes, functions, etc.),
\textit{language} (language in which the dataset's subjects are written), 
\textit{labels} (method to generate labels for the training data),
\textit{granularity} (unit of the code to be classified, e.g., classes, files, lines, or tokens),
\textit{ML technique} (technique or class of neural network being used),
\textit{vulnerability types} (kinds of vulnerabilities being detected),
and
\textit{scope and applicability} (intra- or inter-project).
Many approaches rely on a rough granularity, classifying either whole programs~\cite{Grieco.2016}, files~\cite{Shin.2010}, components~\cite{Neuhaus.2007}, or functions~\cite{Yamaguchi.2011}, which makes it impossible to pin down the precise location of a vulnerability. A more fine-grained granularity would be at the level of multiple lines of codes~\cite{Li.2018,Russell.2018}.
In the following, we discuss approaches that use ML and then particularly deep learning.

\subsubsection{Machine learning for vulnerability detection}

Morrison et al.~\cite{Morrison.2015} examine security vulnerabilities in Windows~7 and~8 with various ML techniques (logistic regression, naive-Bayes, support vector machines and random forest classifiers), achieving a very low precision and recall.

Pang et al.~\cite{Pang.2015} use labels from an online database and aim for classifying whole Java classes as vulnerable or not.
Working on a small dataset of four Java Android applications, they apply an n-gram model in combination with feature selection (ranking) to combine related features and reduce the number of irrelevant features to be considered. Afterwards, they use support vector machines for the sake of learning.
While prediction accuracy within the same project is promising, lower performance has been achieved for cross-project prediction. 

Shar et al.~\cite{Shar.2013b} apply ML to reduce false positives in detecting XSS and SQLI vulnerabilities in PHP code. They first select some code attributes manually and then train a multi-layer perceptron to complement static analysis tools. Compared to static analysis, they detected fewer vulnerabilities, but also achieved lower false positive rates.
In their later work~\cite{Shar.2013}, they use a hybrid approach including dynamic analysis, improving their previous results  on six PHP projects. They also experiment with unsupervised predictors which are, however, less accurate. 

Hovsepyan et al.~\cite{Hovsepyan.2012} use a Java Android email client and analyze its source code like natural language, processing files as a whole. After filtering out comments, files are transformed into feature vectors made up from Java tokens with their respective counts in the file (in a bag-of-words-style approach). These feature vectors are classified in a binary scheme as vulnerable or not. A support vector machine is trained to predict whether a file is vulnerable or not. The promising results illustrate that much insight can be gained by just taking source code as natural text. 
In later work, they investigate various classifiers (decision trees, k-nearest-neighbor, naive-Bayes, random forest and support vector machines)~\cite{Scandariato.2014}.
However, their work is limited by the application on a single software repository. 
	
\subsubsection{Deep learning for vulnerability detection}
\label{deep-vulnerability-prediction}

Deep learning has been applied to learn features for fault prediction~\cite{Wang.2016} as well as for vulnerability detection, which is discussed in the following section.

Russell et al.~\cite{Russell.2018} scrape a large codebase of C projects from GitHub, Debian, and synthetic examples from the SATE IV Juliet test suite, collecting a dataset of over 12 million functions. They use three different static analysis tools to generate the binary labels ``vulnerable'' 
and ``not vulnerable'' for the functions, and a randomly initialized one-hot embedding for lexing.
Convolutional neural networks (CNN) and recurrent neural networks (RNN) are explored for feature extraction, followed by a random forest classifier as the neural networks did not perform well on classification on their own. The CNN performed best, allowing for fine-tuning of precision and recall against each other.
This work is among the first to use deep learning directly on source code from a \textit{large} codebase. Moreover, it is able to use a convolutional feature activation map to highlight the suspicious parts in the code, instead of just classifying a whole function as vulnerable.
	
Liu et al.~\cite{Liu.2018} assume that violations being routinely fixed are true positives, otherwise they are likely to be unimportant or false positives.
Using Findbugs, they investigate revisions from 730 Java projects to identify fixed and non-fixed violations in order to collect code patterns defining actual violations.
These code patterns are encoded into a vector space using word2vec, the discriminating features are learned with a CNN, and an X-means clustering algorithm is used to group violations with learned features.
An interesting result is that a chunk of just 10 lines of code or less is sufficient to capture the relevant context for 90\% of the violations.
Furthermore, their results show that security-related violations are relatively rare (0.5\% of violation occurrences), while they are widespread across 30\% of the projects. Only a small fraction of violations is ever fixed.
The CNN yields patterns that are largely consistent with the violation description provided by Findbugs and that are used to generate fix patterns.
Roughly one third of a test set of violations can be fixed with one of the top-five fix patterns. 
Moreover, ten open-source Java projects were analyzed to make change requests to their developers based on generated fixes. Out of the 116 requests, 67 have been accepted. However, their tool can only suggest patches that correspond to fix patterns previously found in the database.

Although Gupta et al.~\cite{Gupta.2017b} and Dam et al.~\cite{Dam.2016b} have shown that Long Short Term Memory Networks (LSTM) are highly suitable for modeling source code and fixing errors in C code, the latter were the first to use such networks to automatically learn features for predicting security vulnerabilities~\cite{Dam.2017}. They take a publicly available dataset of 18 Java applications and extract the code of all methods 
using abstract syntax trees (ASTs) and replacing some tokens with generic versions. LSTMs are then used for training syntactic and semantic features for a random forest classifier. 
	
VulDeePecker is a deep-learning based vulnerability detection system~\cite{Li.2018}.
The authors also present the first dataset of vulnerabilities intended for deep learning approaches, which stems from popular C and C++ open-source products derived from the National Vulnerability Database and the Software Assurance Reference Dataset. 
VulDeePecker is designed as a tool that does not rely on humans to define features and still provides a satisfyingly low rate of both false negatives and false positives.
Files are split into so-called code-gadgets, semantically related lines of code that are grouped together, and the focus is on key points of library and API calls. 
However, only two different kinds of vulnerabilities are considered: buffer errors and resource management errors. 

Harer et al.~\cite{Harer.2018} train an LSTM to detect and fix vulnerabilities in the synthetic SATE IV codebase of C vulnerabilities. They were able to leverage a sequence-to-sequence approach to produce fixes for found vulnerabilities, although it is hard to measure and compare their success. Similarly, although they are not focusing on security vulnerabilities, Gupta et al.~\cite{Gupta.2017} use RNNs in a sequence-to-sequence setup to fix buggy C code.

\subsubsection{Positioning of \VUDENC}
In contrast to the work by Li et al.~\cite{Li.2018}, Pang et al.~\cite{Pang.2015}, Hovsepyan et al.~\cite{Hovsepyan.2012} and Dam et al.~\cite{Dam.2017}, \VUDENC uses a wide codebase and not only a selected number of projects. The predictions are not only applicable within the same file or the same project, but generalize to any other source code.
While the other approaches classify whole files or, in the case of Li et al.~\cite{Li.2018}, take only API and function calls into account, \VUDENC differs from them by choosing a fine granularity. It is more comparable to the work of Russell et al.~\cite{Russell.2018} and Ma et al.~\cite{Ma.2017}, as vulnerabilities are not just detected at the file level, but at specific positions within the code, which is presumably more useful for developers. \VUDENC further highlights different tokens in different colors depending on the confidence level of the classification.

Similarly to Hovsepyan et al.~\cite{Hovsepyan.2012} but in contrast to Ma et al.~\cite{Ma.2017}, Yamaguchi et al.~\cite{Yamaguchi.2012} and Liu et al.~\cite{Liu.2018}, \VUDENC does not transform source code into a structure like an AST, but takes it as plain text. It follows the natural language approach and aims to use as little assumptions as possible, leaving the extraction of features from the source code entirely to the trained model.
In \VUDENC, the labels for the dataset are not generated by a static analysis tool as in~\cite{Russell.2018,Dam.2017,Hovsepyan.2012}.
The idea of \VUDENC is to be independent of manually designed features that are the main limitation of static analysis tools.
However, the goal is not to model an existing static analysis tool, but to learn features without initial assumptions. Therefore, \VUDENC relies on a similar assumption as Liu et al.~\cite{Liu.2018}, namely that code that was patched or fixed has a high chance of having been vulnerable before the fix. Accordingly, the labeling is based on GitHub commits, which allows the discovery of vulnerability patterns that have not yet been manually included in static analysis tools.

In \VUDENC, the dataset used for training consists of natural code from real-life software projects, as opposed to synthetic databases designed to provide clear examples for vulnerabilities. 
In this aspect, \VUDENC differs from the approaches of Russell et al.~\cite{Russell.2018} and Li et al.~\cite{Li.2018}. Moreover, this aspect makes \VUDENC agnostic towards specific projects with their own characteristics, and therefore to some extent robust against the threats to validity that would result from a more narrow approach.

The machine learning model used by \VUDENC is an LSTM, the same kind of network as used by Li et al.~\cite{Li.2018} and Dam et al.~\cite{Dam.2017}. In comparison, the architecture and preprocessing of data is less complex in \VUDENC. Many other approaches use either different deep learning models (e.g., CNNs and RNNs in~\cite{Russell.2018}), or entirely different ML approaches (e.g., support vector machines in~\cite{Pang.2015}).

Finally, the focus of \VUDENC is on detecting vulnerabilities in Python code, as opposed to other approaches that are concerned with Java, C, C++, or PHP. We found no other approach that uses similar techniques and targets Python code as \VUDENC does.

In conclusion, \VUDENC is distinct from previous research in many ways by expanding the work in the area of vulnerable code pattern analysis. A large dataset of source code written in Python is collected from GitHub, filtered, preprocessed, and labeled based on the information from commits. Several different types of vulnerabilities are taken into consideration, and source code from many different projects is collected. The resulting dataset of natural code containing vulnerabilities is made available for further research. Samples are generated by dividing the code in overlapping snippets that capture the immediate context of some tokens. The samples are embedded in numerical vectors using word2vec. An LSTM network is trained to extract features, and applied to classify code that was not used in training, highlighting the exact locations within the code that are potentially vulnerable.

\section{Approach and Methodology}
\label{sec:approach}

\begin{figure*}[ht!]
	\centering
	\includegraphics[width=\linewidth]{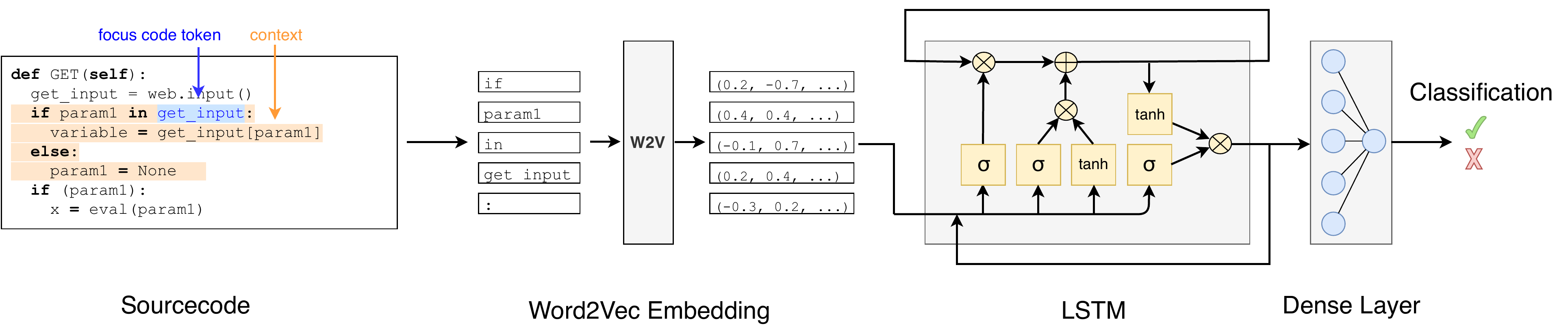}
	\caption{Overview of the \VUDENC approach.}
	\label{fig:architecture}
\end{figure*}

\VUDENC (see Figure \ref{fig:architecture}) works on code tokens and their surrounding tokens to understand the context in which a token appears. The code is embedded in numerical vectors using a word2vec model. Afterwards, an LSTM network is used to recognize features of vulnerable code, and to classify code as \textit{vulnerable} or \textit{not vulnerable} using a final dense layer with a single output neuron. In the following, we will document the specific design choices behind \VUDENC.

\subsection{Choosing a programming language}

Previous works have mostly been focusing on statically-typed languages like Java, C and C++~\cite{Bellon.2007,Russell.2018,Liu.2018,Dam.2017, Rolim.2018}. However, the programming language Python has not received such attention. According to several online rankings, Python is one of the most important and popular programming languages~\cite{AyeshaCuthbert.15.4.2019, VidushiDwivedi.}, and it is the third most used language on GitHub after Javascript and Java~\cite{Github.com.19}. Although Python is known for its usefulness in data science and statistics, this is not its only area of application. With popular web frameworks such as Django and Flask, Python is used to create dynamic websites and web apps, and is therefore also subject to the wide range of security problems occurring in web-based systems. To fill this gap, \VUDENC focuses on source code written in Python.

\subsection{Choosing vulnerability types}
To cover common vulnerabilities, lists of typical security issues such as OWASP Top 10~\cite{OWASPFoundation.} are taken into account. Furthermore, related work~\cite{Zhou.2017,Medeiros.2014,Yamaguchi.2012} provides a starting point for types of vulnerabilities that should be included. Some vulnerability types are excluded because there were only 50 or fewer distinct commits on GitHub found that relate to them. Some other keywords, such as "hijacking", were used figuratively a lot, or were used predominantly in code of offensive applications. After excluding those, a final set of seven typical and widespread vulnerability types are chosen: SQL injection, cross-site scripting, command injection, cross-site request forgery, path disclosure, remote code execution, and open redirect vulnerabilities. For each vulnerability type, a distinct dataset is collected and a separate model is trained to recognize them in source code.
	
\subsection{Data source and labeling} \label{subsec:labeling}
In previous approaches, researchers were able to get better results in predicting vulnerabilities if they applied their model to code from the same project as it has been trained on~\cite{Pang.2015,Dam.2017}. Using a model trained on one project to find vulnerabilities in a different project (cross-project prediction) resulted in a sharp decrease in precision and recall.
Furthermore, the best results were achieved when working on a (partially) synthetic dataset, as opposed to code from ``real'' projects~\cite{Russell.2018,Li.2018}.
With \VUDENC, we aim to make use of a large dataset of real-life source code and train a model that can be applied to any code, not restricted to a single project.

The full dataset is gathered from projects publicly available on GitHub, for several reasons: First, GitHub is the largest host of source code in the world, so it is unlikely that the amount of available useful data will be too little for this application. Second, in contrast to synthetic codebases, nearly all projects on GitHub contain ``natural'' source code in the sense that they are actual projects used in practice. Third, the data is public, making it easier to re-examine and replicate the work, which is not easy for other work that focuses, for instance, on proprietary code.

Since GitHub is also a version control system, it is centered around commits, and as Zhou et al.~\cite{Zhou.2017} suggest, it is practical to look at commits to detect vulnerabilities. A commit that fixes a bug or vulnerability can be described as a patch, consisting of a pair of software versions, a buggy one and an updated and (hopefully) correct one. By analyzing the differences between the old and the new version, vulnerable code patterns can be learned.

Similarly to Li et al.~\cite{Li.2018}, the context of each piece of code is taken into account for labeling. Initially, the data is collected in the form of commits that contain security-related fixes. The code sections that were changed or deleted in such a commit can be labeled as ``vulnerable'', as well as the code in the closest proximity to it (which accounts for the context). The remaining code, as well as the new version after the fix, is labeled as ``(probably) not vulnerable''.
	
To correctly process the data and to avoid an unbalanced dataset~\cite{TanTDM15}, vulnerable and not vulnerable parts must be treated alike and proportional in the labeling step. The idea is to split the data into blocks, and blocks are labeled as vulnerable if they overlap with a vulnerable code segment, otherwise, they are labeled as clean.
	
\begin{figure}[ht]
	\centering
	\includegraphics[width=1.0\linewidth]{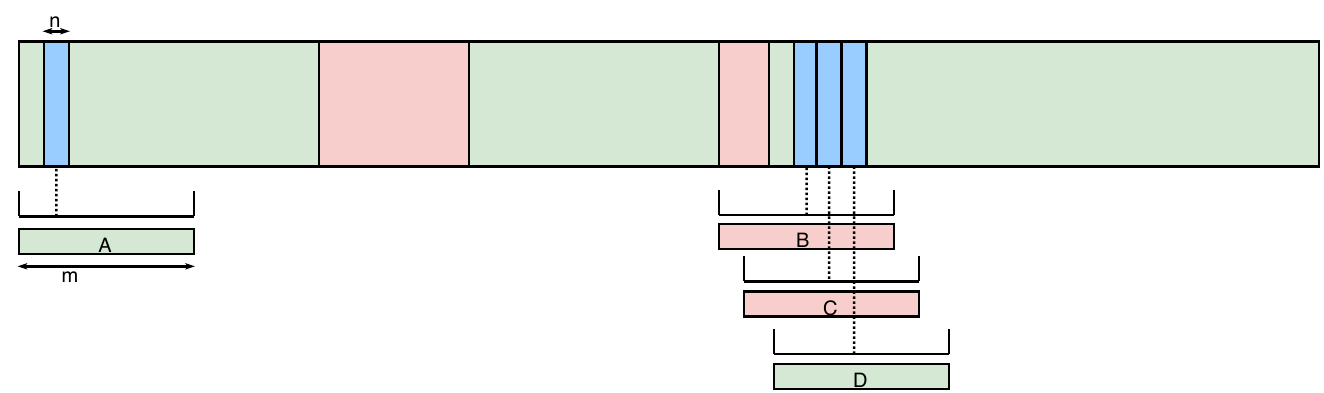}
	\caption{Process of splitting the whole code with vulnerable (red) and non-vulnerable (green) parts in snippets for the dataset.}
	\label{fig:FocusBlocks}
\end{figure}

The labeling procedure works as follows.
Similarly to the work of Hovsepyan et al.~\cite{Hovsepyan.2012}, code comments are filtered out, as they are unlikely to cause a vulnerability. A small focus window traverses through the whole source code in steps of length $n$, as can be seen in Figure~\ref{fig:FocusBlocks}. Four positions of the focus window are depicted in blue. The focus window starts and stops at a character that marks the end of a token in Python, for instance, a colon, a bracket or a whitespace, to prevent cutting tokens in half. For the focus window, the surrounding context of roughly length $m$, also starting and stopping at the border of code tokens, is determined, with $m > n$. If the focus window is close to the beginning of the file, the context will mostly lie behind it (e.g., block A), and if it is located in the middle, the surrounding context will be spanning a snippet that lies equally before and after the focus window. This results in a number of overlapping blocks. If the whole block contains partially vulnerable code (e.g., blocks B and C), it is labeled as vulnerable, otherwise it is labeled as clean. This ensures that code snippets containing a vulnerability are marked as such. The parameters $n$ and $m$ are subject to optimization, their ideal values will be determined experimentally. Liu et al.~\cite{Liu.2018} state that, most often, a chunk of 10 lines of code is sufficient to capture the relevant context of a vulnerability.
	
The datasets we created and use are available on Github~\cite{Wartschinski.2.12.2019b} and Zenodo~\cite{Wartschinski.2.12.2019c}. 
The next step is to transform those code blocks, which are just lists of Python tokens, into lists of numerical vectors. For this, the word2vec embedding is needed.
	
\subsection{Representation of source code}
	
\subsubsection{Choosing a representation}
Simple approaches like bag-of-words representations have yielded mediocre results in the past, and are by design not able to capture the semantic context of code. 
Liu et al.~\cite{Liu.2018} argued that an AST representation is necessary to mine patterns from code, while Russel et al.~\cite{Russell.2018} and Hovsepyan et al.~\cite{Hovsepyan.2012} demonstrate that this is not necessarily the case, as code can also be modeled with textual representations. Furthermore, Dam et al.~\cite{Dam.2016} argue that, alongside with human-engineered features and software metrics, ASTs might not be able to capture the semantics hidden deeply in source code.
Code has many similarities with natural language text: repetitiveness of certain structures and common patterns, locality (repetitions occur in a local context), and long-term dependencies~\cite{Dam.2016,Hindle.2012}. Furthermore, code is written by humans, who have a tendency to gravitate towards conventional patterns and repetition of typical structures \cite{Allamanis.2018}.
	
Deep neural networks are designed for the task of modeling sequential data of various kinds, including natural language, sensor data and even music~\cite{Kumar.2019}, and have yielded outstanding results in these areas. Therefore, they have been applied successfully to modeling source code directly as text~\cite{Russell.2018,Gupta.2017,Li.2018,Harer.2018}.
Taking all this into account, our choice is to work directly on source code as text, without reliance on ASTs. Since snippets of code are taken as samples, the approach could be called a version of an n-gram approach, although the length of the snippets is much longer than usually in n-grams (covering not only a handful of tokens, but several lines of code). To account for the locality property of code~\cite{Tu.2014}, emphasis will be placed on the context around each code token for feature learning.

\begin{figure*}[ht!]
\centering
\includegraphics[width=0.9\linewidth]{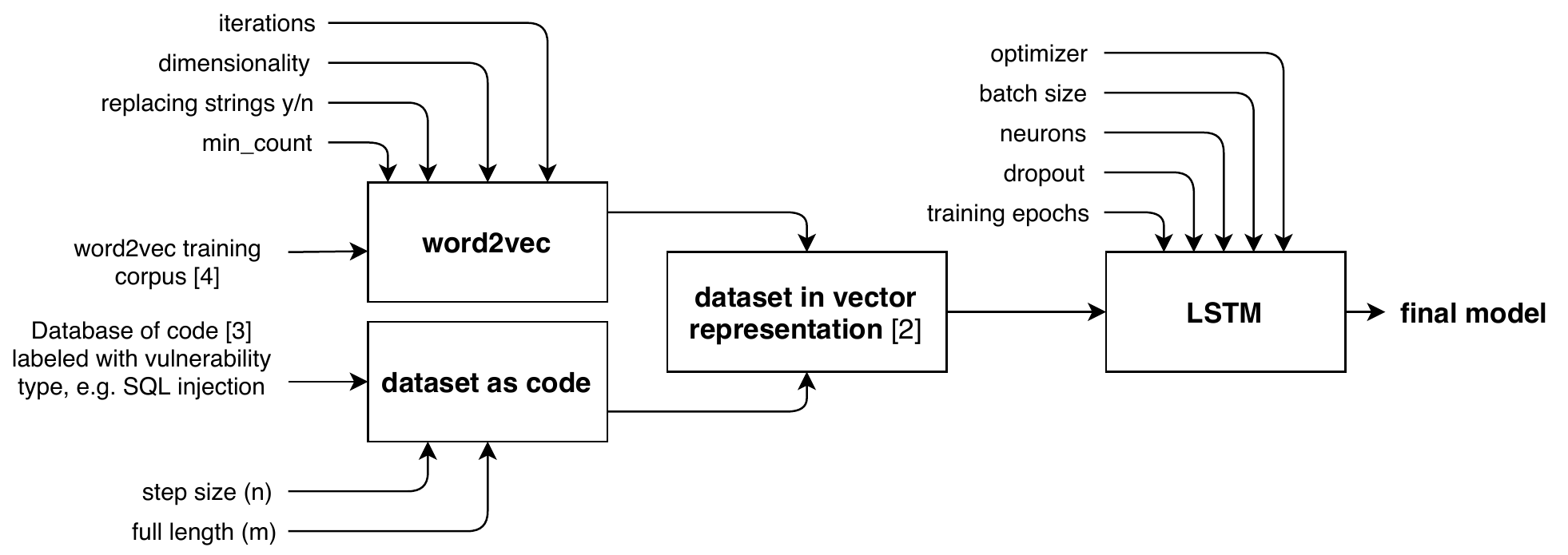}
\caption{Overview of \VUDENC's (hyper)parameters.}
\label{fig:hyperparameters}
\end{figure*}

\subsubsection{Choosing granularity}
As Morrison  et al.~\cite{Morrison.2015} describe, binary-level predictions and analysis on the level of whole files provide little insight, as developers often already know which files might be sensitive to security vulnerabilities, and developers strongly prefer a much finer approach, if possible at the level of lines of code or instructions.
Dam et al.~\cite{Dam.2017} provide some convincing examples at the beginning of their work, arguing that there exist files with similar metrics, similar structure and even nearly the same tokens, of which one might be clean and another might be vulnerable, despite the similar metrics. A top-down perspective, looking at whole files, can therefore not be as promising as an approach that ``zooms in'' to analyze small snippets of code individually.
	
\VUDENC implements an approach of fine granularity, looking at each token in the code as well as its context. This makes it possible to pin down the exact location of the vulnerability.
	
\subsubsection{Preprocessing the code}
Neural networks work on numerical vectors of uniform size, and therefore it is necessary to represent code tokens as vectors that retain the semantic and syntactic information embodied in the code. In addition, the variables of the vector have to be chosen in such a way that the vectors are manageable in size. A word2vec embedding represents semantically similar code tokens with vectors of high cosine similarity. Word2vec has been successfully used for similar projects before~\cite{Liu.2018}. In addition to its ability to capture the semantic content of tokens, it also requires much smaller vector sizes than, for example, a simple one-hot encoding, which makes it less computationally expensive. It is chosen as the appropriate embedding method for \VUDENC.

\VUDENC works on the full source code, only stripping out the comments, but otherwise leaving the code exactly as it is. To transform a snippet of code in a numerical representation, it is first split up to a list of tokens (operators, identifiers, etc.) by using the ``tokenizer module''\footnote{\url{https://docs.python.org/3/library/tokenize.html}} comprised by the Python standard library. The module provides a dedicated lexical scanner for Python source code. For instance, this module splits a line such as ``def my\_function():'' into five parts: ``def'', ``my\_function'', ``(``, ``)'', and ``:''. Each of those tokens has to be embedded, in other words, represented by a numeric vector. A full section of code is therefore transformed into a vector of vectors of numbers.
	
Since there is currently no pre-trained language model for Python code available, the word2vec model first has to be trained. For this purpose, a corpus of high-quality Python code is acquired, for which we use GitHub. On this corpus, the word2vec model is trained to prepare it for the task to encode Python code tokens as vectors.
The hyperparameters of the word2vec model are (cf. Figure \ref{fig:hyperparameters}):
\begin{itemize}[noitemsep]
\item training iterations: between one and more than hundred
\item vector dimensionality: between 5 and 300
\item replacing string with generic tokens, or leaving them as they are
\item minimum count: between 10 and 5000
\end{itemize}
	
\subsection{Selecting the machine learning model}
The goal of \VUDENC is to create a model that can learn vulnerability features from sequences of code tokens. Source code is, by its very nature, sequential data, as every statement's effect depends on surrounding instructions. To detect a vulnerability, it is not enough to learn that a single token is ``bad'', as this will lead to many false positives. Rather, the goal is to learn that a token is ``bad'' \textit{when used in a specific way}, that is, in combination with surrounding tokens.

Deep learning-based models, especially RNNs and LSTMs, are well-suited to represent the locality of code, while at the same time being able to capture much longer contexts than n-grams. They are designed for exactly the kind of task required in \VUDENC and have been successfully used in modeling code~\cite{Dam.2016}. Therefore, an LSTM is chosen as the model for this work.
The hyperparameters of the LSTM are (cf. Figure \ref{fig:hyperparameters}):
\begin{itemize}[noitemsep]
\item optimizer: (see next Section \ref{optimizer}).
\item batch size: number of samples which are shown to the network to be processed before the weights are updated again. Batch sizes of 32, 64 and 128 samples are typically used.
\item number of neurons (or units): more neurons allow the model to learn a more complex structure, but also require a longer time to train the model.
\item dropout: input and recurrent connections to LSTM units are sometimes randomly excluded from the next step of the training, which is a regularization method to reduce the chance of overfitting. A typical dropout is chosen between 10\% and 50\%.
\item number of training epochs: the number of times that the learning algorithm will work through the whole training dataset. Typical numbers for epochs in the literature include 10, 100, 500 or even 1000 epochs.
\end{itemize}
	
\subsection{Choosing the optimizer}\label{optimizer}
The goal of \VUDENC is to achieve a high precision and recall, therefore the F1 score (see Equation \ref{eq:F1}) is chosen as the loss function that needs to be minimized by the optimizer. The optimizer is part of \VUDENC's parameter set (cf. Figure \ref{fig:hyperparameters}). The most basic optimizer, stochastic gradient descent, can not be applied in this case, since the F1 loss is not necessarily a convex function. The adam optimizer (named after the technique of ``adaptive moment estimation'') chooses a learning rate dynamically. It was published in 2014~\cite{Kingma.2014} and has been designed specifically for deep neural networks, where it achieves good results fast and is often taken as a go-to optimization algorithm for many problems. Looking at other works in the field, Li et al.~\cite{Li.2018} use adamax, Russell et al.~\cite{Russell.2018} leverage the standard adam optimizer and Dam et al.~\cite{Dam.2017} use RMSprop, although the applicability depends strongly on the specifics of the dataset. For \VUDENC, the adam optimizer will be be used as a starting point which can be compared empirically to other optimizers (adagrad, adamax, nadam and RMSprop) to review which yields the best results in practice.

\subsection{Presenting the results}\label{results}	
Finally, the results from the trained classifier need to be used to visualize vulnerabilities in  the source code. The code is split into blocks in exactly the same way as before (using a small focus area and a sliding context window as described in Section~\ref{subsec:labeling}). The focus area traverses through the code, and for each new step the surrounding context of a few tokens is taken, the model makes a prediction based on that context as input, and the prediction is used as the vulnerability classification for the focus area. 

\begin{figure}[ht!]
\begin{subfigure}{0.5\textwidth}
		\centering
		\includegraphics[width=1.0\linewidth]{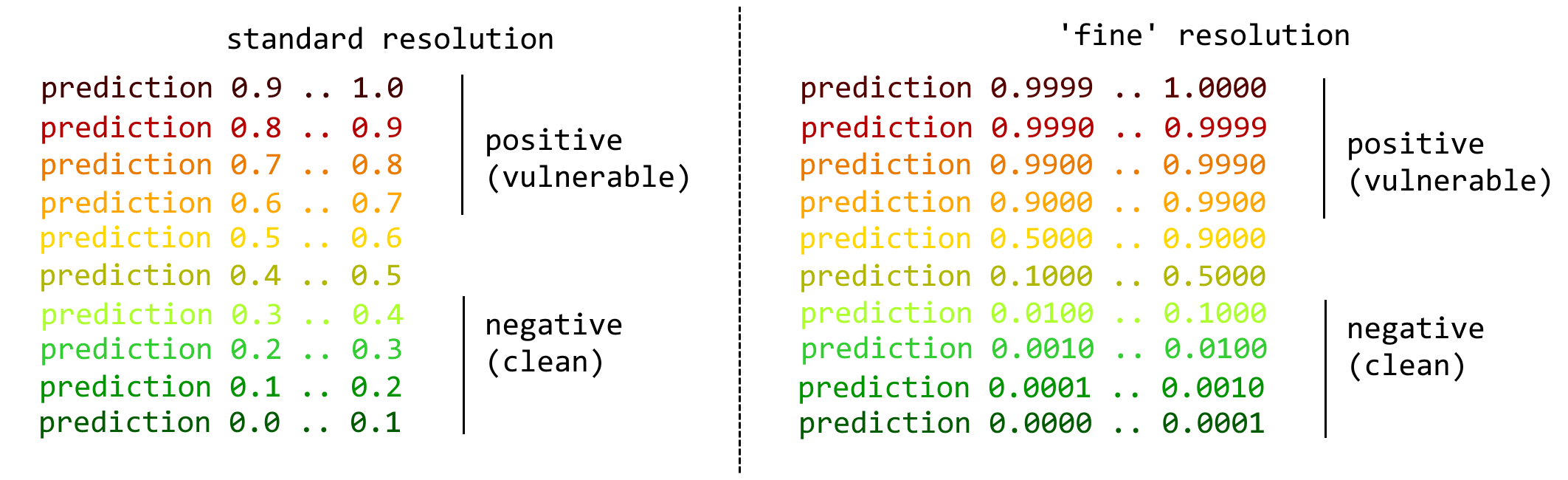}
		\caption{Colors and confidence levels.}
		\label{fig:colors}
\end{subfigure}
\vspace{2mm}
\begin{subfigure}{0.5\textwidth}
		\includegraphics[width=0.95\linewidth]{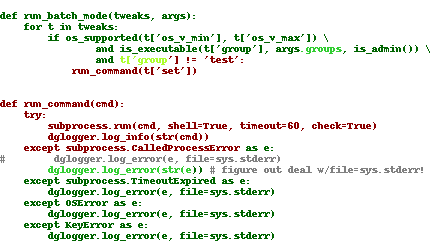}
		\caption{Example of a command injection vulnerability (fine resolution).}
		\label{fig:example}
\end{subfigure}
\caption{Visualization of detected vulnerabilities in the source code.}
\end{figure}

Colors (see Figure \ref{fig:colors}) are used to highlight the different confidence levels of the classification. Figure \ref{fig:example} presents an example of a result for a code snippet that contains a command injection vulnerability. The red parts are where the vulnerability is suspected to be located. The subprocess routine is called with the parameter "shell=True", which means that all input is interpreted according to the syntax rules of the invoked shell, including characters such as ";" or "$|$", which may result in arbitrary program execution. Of course, this can only happen if the user can influence the variable "cmd" somehow (which was actually the case in this example). Setting the shell parameter to "true" is often not necessary but rather dangerous and should be avoided. Therefore, it is desirable that the code fragment is marked red to indicate that it might contain a vulnerability.	

\section{Empirical Evaluation}\label{sec:evaluation}
In this section, we present the evaluation of \VUDENC.
Our tool and all evaluation data are publicly available on Github~\cite{Wartschinski.2.12.2019b} and Zenodo~\cite{Wartschinski.1.12.2019,Wartschinski.2.12.2019, Wartschinski.2.12.2019c} respectively to follow open-science principles and enable the replication of our experiments.

\subsection{Research questions}
In our evaluation we investigate the following four research questions: \\

\noindent
{\setlength\tabcolsep{2pt}
\begin{tabu}{ll}
\textbf{RQ-1:} & \textit{Can a data set of suitable Python source code be} \\
& \textit{mined from Github for common vulnerability types?} \\
\textbf{RQ-2}: & \textit{Is the word2vec model effective, and how do the} \\
& \textit{hyperparameters influence the overall results?} \\
\textbf{RQ-3}: & \textit{Is the LSTM model effective, and how do the} \\
& \textit{hyperparameters influence the overall results?} \\
\textbf{RQ-4}: & \textit{How effective is VUDENC in detecting vulnerabili-} \\
& \textit{ties as measured with precision, recall and F1?}
\end{tabu}
}

\subsection{Evaluation setup}
\subsubsection{Infrastructure}
The model training and all experiments were conducted on a machine running openSUSE Leap 15.1 equipped with 8 Quad-Core-AMD 8384 2.7 GHz and 64 GB of memory.

\subsubsection{The baseline model}\label{baseline}
To show the effects of various hyperparameter changes and tweaks, a baseline model was created.
Its hyperparameters are not optimal, but it can be used to demonstrate how other hyperparameters cause better or worse results, since some configuration has to be taken as a starting point.
The baseline model has a focus area step size $n$ of 5, a context length $m$ of 200, and works on the data set for SQL injections.
It has 30 neurons and is trained for 10 epochs with a dropout and recurrent dropout of 20\%, with a batch size of 200, using the adam optimizer.
Although more epochs would almost certainly lead to better outcomes, we had to limit the number of epochs, since many combinations have to be tested, which is a costly process.

\subsection{Evaluation metrics}\label{Evaluation}
For the purpose of prediction and classification, four key concepts are usually the basis for evaluation: true positives (TP), true negatives (TN), false positive (FP) and false negatives (FN). Positive and negative refer to the prediction, meaning that a prediction of ``vulnerable'' would be a positive and a prediction of ``not vulnerable'' would be a negative. The terms true and false refer to whether the prediction corresponds to the actual value or external judgment. 
	
The \textbf{precision} (cf.~Equation~\ref{eq:precision}) is the rate of true positives within all positives. It measures how precise the model is in terms of how many of the predicted positives are actual positives, or phrased differently, how much trust can be placed in the classification of a positive and how many false alarms are produced. A positive is interpreted as true if it has an overlap with a known vulnerability, which is justifiable since the snippets are very short. The \textbf{recall} (cf.~Equation~\ref{eq:recall}), also called sensitivity, is a measurement for the rate of positives that were correctly identified in comparison to the total number of actual positives. The \textbf{F1 score} (cf.~Equation~\ref{eq:F1}) is a balanced score (harmonic mean) that takes both precision and recall into account.

\begin{equation}
	Precision = \frac{TP}{TP + FP}
\label{eq:precision}
\end{equation}
\begin{equation}
	Recall = \frac{TP}{TP + FN}
\label{eq:recall}
\end{equation}
\begin{equation}
	F1 = 2 \cdot \frac{Precision \cdot Recall}{Precision + Recall}
\label{eq:F1}
\end{equation}

Ideally, the model would achieve a near 0\% rate for false positives and false negatives, meaning that precision and recall both are close to 1, as well as the F1 score.	

Finally, another metric is the \textbf{accuracy} (cf.~Equation~\ref{eq:accuracy}), which is the fraction of correct predictions compared to all predictions. The accuracy is defined as follows:
\begin{equation}
    Accuracy = \frac{TP + TN}{TP + FP + TN + FN}
\label{eq:accuracy}
\end{equation}
However, accuracy does not provide much insight when there is a class imbalanced data set, meaning that there are many more positives than negatives or vice versa. In the case of vulnerability detection, it is indeed the case that most code fragments will be clean and vulnerabilities are relatively rare. As a result, we will use precision, recall, and the F1 score to evaluate the models. The accuracy is just reported for completeness reasons.

\begin{table*}[ht]
\begin{small}
\caption{Mined data sets from GitHub.}
\label{tab:subjects}
\centering
\begin{tabu}{|l|r|r|r|r|r|r|r|}
\hline 	
\multirow{2}[2]{*}{\textbf{Vulnerability type}} &
\multirow{2}[2]{*}{\centering\textbf{\# repositories}} &
\multirow{2}[2]{*}{\centering\textbf{\# commits}} &
\multirow{2}[2]{*}{\centering\textbf{\# changed files}} &
\multirow{2}[2]{*}{\centering\textbf{LOC}} &
\multirow{2}[2]{*}{\centering\textbf{\# sep.\ functions}} &
\multirow{2}[2]{*}{\centering\textbf{\# chars}} \\
&&&&&& \\
\hline 			
SQL injection~\cite{OWASP_SQLInjection} & 336 & 406 & 657 & 83,558 & 5,388 & 3,960,074 \\
XSS~\cite{OWASP_XXS} & 39 & 69 & 81 & 14,916 & 783 & 736,567 \\
Command injection~\cite{OWASP_CommandInjection} & 85 & 106 & 197 & 36,031 & 2,161 & 1,740,339 \\
XSRF~\cite{OWASP_XSRF} & 88 & 141 & 296 & 56,198 & 4,418 & 2,682,206 \\
Remote code execution~\cite{OWASP_CodeInjection} & 50 & 54 & 131 & 30,591 & 2,592 & 1,455,087 \\
Path disclosure~\cite{OWASP_CommandInjection} & 133 & 140 & 232 & 42,303 & 2,968 & 2,014,413 \\
Open redirect~\cite{CWE_OpenRedirect} & 81 & 93 & 182 & 26,521 & 1,762 & 1,295,748 \\
\hline
\end{tabu}
\end{small}
\end{table*}

\subsection{RQ-1 Mining data set}
We collected a large data set, which (before filtering) consists of 25,040 vulnerability-fixing commits in 14,686 different repositories from GitHub. Commits that changed the code in too many places were filtered out, in order to ensure that the changed code is actually related to the commit message. While there are still occasionally other changes in the commits, those do not introduce a systematic error.
For each vulnerability type we created specialized data sets.
Only files that are Python source code are part of the data sets.
Excessive duplicates are excluded as well as projects that are showcases for vulnerabilities or demonstrate hacking techniques, and low-quality or undesired data is filtered out by removing files that are more than 10,000 characters long, contain lots of HTML code, or contain content in their commit message or repository name that suggests that the project is just showcasing or demonstrating a vulnerability (instead of fixing it).

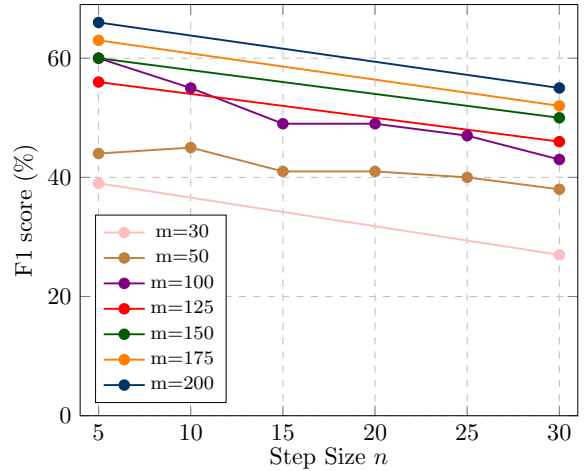
\begin{figure}[h]
\begin{center}
\begin{tikzpicture}[scale=\plotscale]
\begin{axis}[
  xlabel=Step Size $n$,
  ylabel=F1 score (\%),
  xmajorgrids=true,  
  ymajorgrids=true,
  grid style=dashed,
  xmin=4, xmax=31,
  ymin=0, ymax=69,
  x label style={at={(axis description cs:0.5,0.03)}},
  y label style={at={(axis description cs:0.06,0.5)}},
  width = \columnwidth,
  legend style={font=\footnotesize,at={(0.03,0.26)},anchor=west},
   ]
\addplot[color=pink,mark=*, thick] table [y=m30, x=n, col sep=comma]{plots/01_dataset_parameter.dat};
\addlegendentry{m=30}
\addplot[color=brown,mark=*, thick] table [y=m50, x=n, col sep=comma]{plots/01_dataset_parameter.dat};
\addlegendentry{m=50}
\addplot[color=violet,mark=*, thick] table [y=m100, x=n, col sep=comma]{plots/01_dataset_parameter.dat};
\addlegendentry{m=100}
\addplot[color=red,mark=*, thick] table [y=m125, x=n, col sep=comma]{plots/01_dataset_parameter.dat};
\addlegendentry{m=125}
\addplot[color=darkgreen,mark=*, thick] table [y=m150, x=n, col sep=comma]{plots/01_dataset_parameter.dat};
\addlegendentry{m=150}
\addplot[color=orange,mark=*, thick] table [y=m175, x=n, col sep=comma]{plots/01_dataset_parameter.dat};
\addlegendentry{m=175}
\addplot[color=darkblue,mark=*, thick] table [y=m200, x=n, col sep=comma]{plots/01_dataset_parameter.dat};
\addlegendentry{m=200}
\end{axis}
\end{tikzpicture}
\caption{Data set parameters: step size $n$ and sample length $m$.}
\label{fig:mn}
\end{center}
\end{figure}

The data set consists of samples, each of them representing a small snippet of code centered around a single token.
There are two parameters for processing the data with regard to these samples.
(1) $n$, the step size for moving the focus point through the source code, and
(2) $m$ the size of the context window around the token in focus, i.e., the full length of the code sample.
Both are measured in number of characters.
For example, a smaller step size means more samples in total, and more overlap for the samples.
We performed experiments to find well performing values for $n$ and $m$.
All hyperparameters of the LSTM model are the default values described before in Section~\ref{baseline}, and the ideal word2vec model determined before is used.
The results are presented in Figure \ref{fig:mn}.

A larger $n$ consistently leads to worse results.
This is most likely a result of the fact that if the gaps between one focus point and the next one are large, there is not much overlap between their surrounding context, the moving window that makes up the code snippets.
If the focus moves in very small steps, the code snippets have a lot of overlap, and consequently, a single token will appear several times, for instance, at the end of one snippet, at the center of the next one, and at the start of the one after that.
This means that for every vulnerability, there are samples that show the relevant code with more of the context before and after it, possibly making it easier for the model to learn which part is actually the source of the vulnerability.

The model performs better with a larger full length $m$ of the code snippet making up one sample.
A larger $m$ again leads to more overlap.
The disadvantage here is that the prediction might get a little less accurate, as a large snippet around some token might be classified as vulnerable because somewhere else in the snippet of length $m$ is a vulnerable part.
On the other hand, a larger $m$ also has the advantage that more context can be taken into account for a token, which is exactly why the LSTM was chosen in the first place.
For a full length of more than 200, the samples that were already quite numerous got also relatively large in size, exceeding the computational capabilities of the machines.
Therefore, a step length of n=5 and a full context window length of m = 200 were fixed as the parameters for creating the training set.

The final data sets contain the source code and some information about the commit as well as the vulnerable and the clean segments within the code.
The basic information about them is summarized in Table \ref{tab:subjects}.
We report the number of repositories (\# repo.) and commits (\# commits) that are part of the data set, the number of changed files with vulnerabilities (\# changed file), their lines of code (LOC), the number of separate functions (\# separate functions) within, and the total number of characters (\# chars).
That this data set is \textit{suitable} will be demonstrated in the next sections, by utilizing it in training the model.

In total, seven different types of vulnerabilities are part of the data set:
SQL injection, Cross-site scripting (XSS), Command injection, Cross-site request forgery (XSRF), Remote Code Execution, Path disclosure, and Open Redirect.
All those are vulnerabilities that are widespread and in many cases dangerous to applications and systems~\cite{OWASPFoundation.,CVE}.
Each vulnerability data set stems from 39 up to 336 different repositories or projects,
and contains source code between around 14,000 and 83,000 lines of code, spread over 80-650 files comprising between around 700 and 5.000 functions. \\

\noindent
\hspace{2pt}
\fbox{\begin{minipage}{0.93\columnwidth}
\textbf{Answer for RQ-1}:
Yes, we acquired a satisfyingly large data set from GitHub, which covers seven common vulnerability types: SQL injection, Cross-site scripting (XSS), Command injection, Cross-site request forgery (XSRF), Remote Code Execution, Path disclosure, and Open Redirect.
\end{minipage}}

\subsection{RQ-2 word2vec embedding \& hyperparameters}

The training corpus taken from various Python repositories contains 69,517,343 individual tokens.
The following hyperparameters get evaluated: \textit{vector dimensionality}, \textit{minimum count}, \textit{training iterations}, as well as the outcome of \textit{replacing strings} with generic string tokens.
As already stated, the baseline model (see Section~\ref{baseline}) is used, so all hyperparameters are chosen according to this default configuration unless specified otherwise.
The approach is, in general terms, to train a word2vec model, which is then used to embed the data and train a LSTM model on it.
The performance of the LSTM model is used to judge the quality of the underlying word2vec embedding since the embedding itself cannot be evaluated.
Its effectiveness is determined by the fact that it can be used in the context it is intended for.

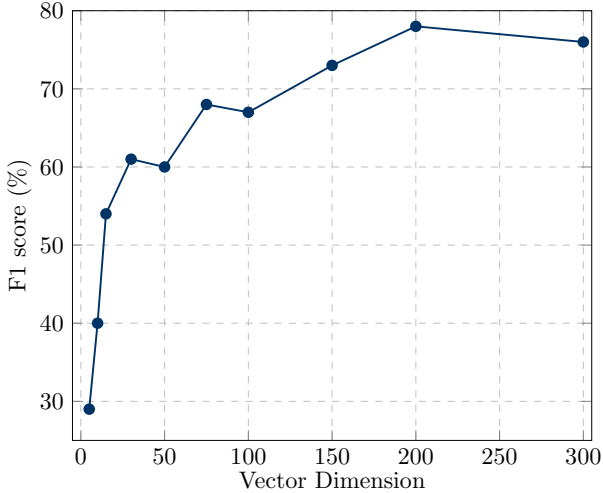
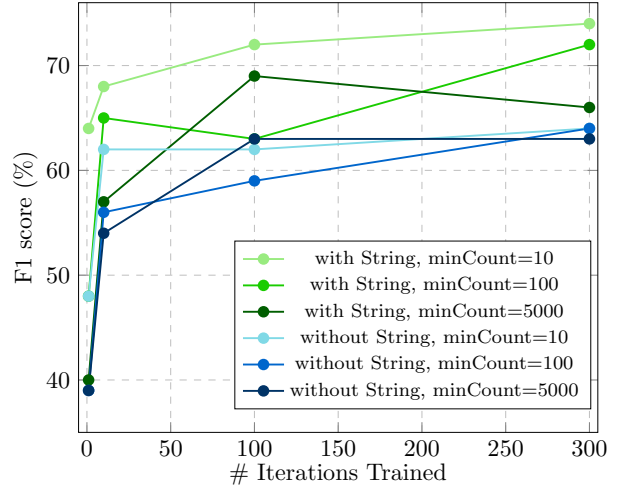
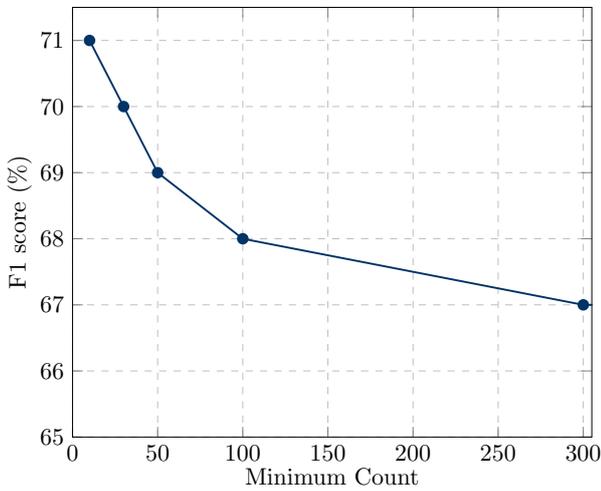
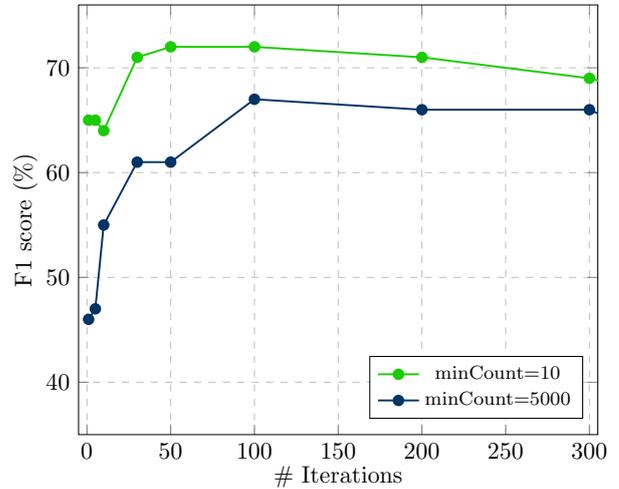
\begin{figure*}[ht!]
\begin{subfigure}{0.5\textwidth}
	\begin{center}
\begin{tikzpicture}[scale=\plotscale]
\begin{axis}[
  xlabel=Vector Dimension,
  ylabel=F1 score (\%),
  xmajorgrids=true,  
  ymajorgrids=true,
  grid style=dashed,
  xmin=-5, xmax=305,
  ymin=25, ymax=80,
  x label style={at={(axis description cs:0.5,0.03)}},
  y label style={at={(axis description cs:0.06,0.5)}},
  width = \columnwidth,
  legend style={font=\footnotesize,at={(0.02,0.25)},anchor=west},
   ]
\addplot[color=darkblue,mark=*, thick] table [y=F1, x=Dimensionality, col sep=comma]{plots/02_dimensionality.dat};
\end{axis}
\end{tikzpicture}
\caption{Results for different vector dimensionalities.}
\label{fig:vector-dimension}
\end{center}
\end{subfigure}
\begin{subfigure}{0.5\textwidth}
	\begin{center}
\begin{tikzpicture}[scale=\plotscale]
\begin{axis}[
  xlabel=\# Iterations Trained,
  ylabel=F1 score (\%),
  xmajorgrids=true,  
  ymajorgrids=true,
  grid style=dashed,
  xmin=-5, xmax=305,
  ymin=35, ymax=76,
  x label style={at={(axis description cs:0.5,0.03)}},
  y label style={at={(axis description cs:0.06,0.5)}},
  width = \columnwidth,
  legend style={font=\footnotesize,at={(0.3,0.25)},anchor=west},
   ]
\addplot[color=lightgreen,mark=*, thick] table [y=withStringMinCount10, x=iterations, col sep=comma]{plots/03_string_replacement.dat};
\addlegendentry{with String, minCount=10}
\addplot[color=green,mark=*, thick] table [y=withStringMinCount100, x=iterations, col sep=comma]{plots/03_string_replacement.dat};
\addlegendentry{with String, minCount=100}
\addplot[color=darkgreen,mark=*, thick] table [y=withStringMinCount5000, x=iterations, col sep=comma]{plots/03_string_replacement.dat};
\addlegendentry{with String, minCount=5000}
\addplot[color=lightblue,mark=*, thick] table [y=withOutStringMinCount10, x=iterations, col sep=comma]{plots/03_string_replacement.dat};
\addlegendentry{without String, minCount=10}
\addplot[color=blue,mark=*, thick] table [y=withOutStringMinCount100, x=iterations, col sep=comma]{plots/03_string_replacement.dat};
\addlegendentry{without String, minCount=100}
\addplot[color=darkblue,mark=*, thick] table [y=withOutStringMinCount5000, x=iterations, col sep=comma]{plots/03_string_replacement.dat};
\addlegendentry{without String, minCount=5000}
\end{axis}
\end{tikzpicture}
\caption{Results for different variants of the string replacement.}
\label{fig:string-replacement}
\end{center}
\end{subfigure}
\begin{subfigure}{0.5\textwidth}
	\begin{center}
\vspace{5mm}
\begin{tikzpicture}[scale=\plotscale]
\begin{axis}[
  xlabel=Minimum Count,
  ylabel=F1 score (\%),
  xmajorgrids=true,  
  ymajorgrids=true,
  grid style=dashed,
  xmin=0, xmax=305,
  ymin=65, ymax=71.5,
  x label style={at={(axis description cs:0.5,0.03)}},
  y label style={at={(axis description cs:0.06,0.5)}},
  width = \columnwidth,
  legend style={font=\footnotesize,at={(0.3,0.25)},anchor=west},
   ]
\addplot[color=darkblue,mark=*, thick] table [y=f1, x=count, col sep=comma]{plots/04_mincount.dat};
\end{axis}
\end{tikzpicture}
\caption{Results for different values as minimum count.}
\label{fig:mincount}
\end{center}
\end{subfigure}
\begin{subfigure}{0.5\textwidth}
	\begin{center}
\vspace{5mm}
\begin{tikzpicture}[scale=\plotscale]
\begin{axis}[
  xlabel=\# Iterations,
  ylabel=F1 score (\%),
  xmajorgrids=true,  
  ymajorgrids=true,
  grid style=dashed,
  xmin=-5, xmax=305,
  ymin=35, ymax=76,
  x label style={at={(axis description cs:0.5,0.03)}},
  y label style={at={(axis description cs:0.06,0.5)}},
  width = \columnwidth,
  legend style={font=\footnotesize,at={(0.56,0.11)},anchor=west},
   ]
\addplot[color=green,mark=*, thick] table [y=minCount10, x=iterations, col sep=comma]{plots/05_iterations.dat};
\addlegendentry{minCount=10}
\addplot[color=darkblue,mark=*, thick] table [y=minCount5000, x=iterations, col sep=comma]{plots/05_iterations.dat};
\addlegendentry{minCount=5000}
\end{axis}
\end{tikzpicture}
\caption{Results for different numbers of iterations.}
\label{fig:iterations}
\end{center}
\end{subfigure}
\caption{Influence of the hyperparameters of the word2vec embedding on the overall results (RQ-2).}
\end{figure*}

\textbf{Vector dimensionality}:
When using word2vec, the code tokens are converted into numerical vectors of a certain length or dimensionality.
The longer those vectors are, the more different \textit{axes} there are for putting words in relation to each other, allowing the word2vec model to capture more complex relationships.
A vector size of less than 100 is unlikely to represent the semantics of Python code well, judging from similar tasks with natural language, where vector sizes of 200 are typical.
To compare different vector lengths, the minimum count of a token to appear in the vocabulary is set to 1,000 and training iterations of the word2vec model are set to 100.
In the experiment, we replace strings with a generic string token.
Using those hyperparameters, the vector dimensionality was varied between 5 and 300.
As is evident from the Figure \ref{fig:vector-dimension}, a reasonable vector size seems to be around 200.

\textbf{String replacement}:
Strings occurring in the Python training file could be replaced with a generic \textit{string} token, as it has been done by other researchers~\cite{Dam.2017,Russell.2018}. Replacing them might reduce the level of detail of the model, but keeping them might put too much focus on the specific content of string tokens---it is difficult to say beforehand what works better.
To compare the two approaches, the length of the embedding vectors is fixed to 200.
The training iterations are varied between 1 and 300, and a minimum count (min\_count) of 10, 100 and 5,000 is compared.

The results are shown in Figure \ref{fig:string-replacement}, where the models that keep strings are marked with green lines, while the models that replace the strings are shown as blue lines.
The versions with the original strings yield consistently better results.
Looking at the data, it can be also already suspected that more iterations and a lower min\_count are beneficial for the overall performance.

\textbf{Minimum count}:
The minimum count defines how often a token has to appear in the training corpus in order to actually get assigned a vector representation.
Tokens that appear less often are simply ignored and will not be encoded (and instead skipped later when the complete lists of tokens are converted to lists of vectors).
This is mostly to ignore rare variable names, strings or other identifiers that are not relevant.
The word2vec model is trained with strings kept as they are, for 100 iterations, and with a vector size of 200.
The min\_count is chosen between 10 and 5000, with the results shown in Figure \ref{fig:mincount} (we omitted the results for min\_count=5000, which are the same as for min\_count=300).
Although it might have seemed reasonable to assume that ignoring rare tokens would improve the performance, this was indeed not the case.
The model performs better when hardly any token is being ignored.

\textbf{Iterations}:
The number of iterations defines the number of repetitions in training the word2vec model.
It is expected that after a certain number of iterations, there will be no additional benefit in further training.
As before, the model is trained on a corpus with original strings included, with a dimensionality of 200, and a min\_count of 10 or 5,000.

The results, presented in Figure \ref{fig:iterations}, show that until 50 to 100 iterations, more iterations lead to a better performance of the model.
Increasing the iterations to 300 does not improve the model performance, but instead reduces it, and also results in much longer training time.
Note that it can again be confirmed that a lower min\_count generally results in better performance.

Our experiments show that the word2vec hyper\-parameters do result in significantly different performances of the LSTM model, spanning a difference of roughly 25 percentage points for the LSTM's F1 score between the best and worst word2vec parameters.
It can therefore be concluded that the careful consideration of the hyperparameter values is worthwhile, and the quality of the embedding has an influence on how well the final model is able to learn features.
Overall, the use of a word2vec embedding can definitely be recommended as a feasible approach for future, similar research. \\

\noindent
\hspace{2pt}
\fbox{\begin{minipage}{0.93\columnwidth}
\textbf{Answer for RQ-2}:
Yes, the word2vec model is suitable as an effective embedding.
Our final word2vec model will encode code tokens in numerical vectors of 200 dimensions, not replace any strings, require a min\_count of 10 for tokens to be included, and will be trained for 100 iterations.
\end{minipage}}

\subsection{RQ-3 LSTM model \& hyperparameters}

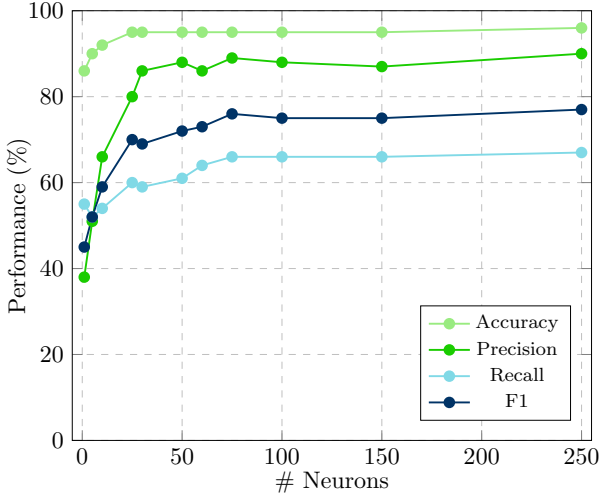
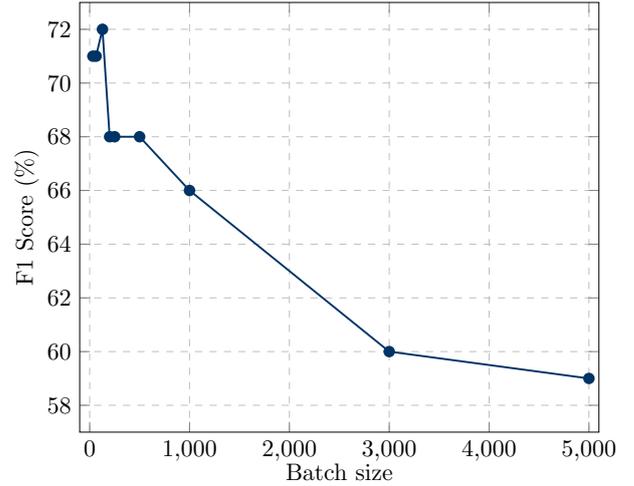
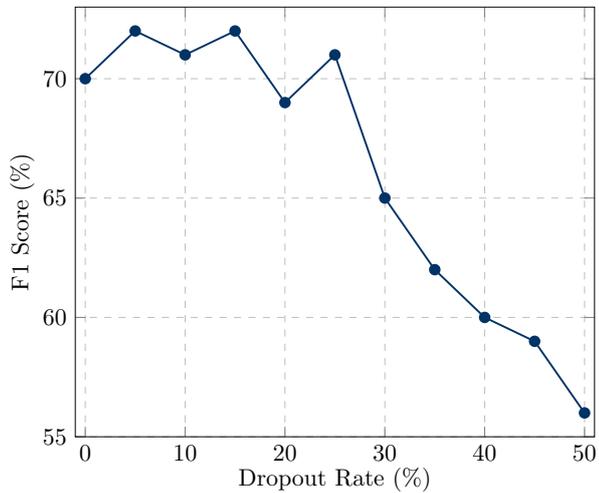
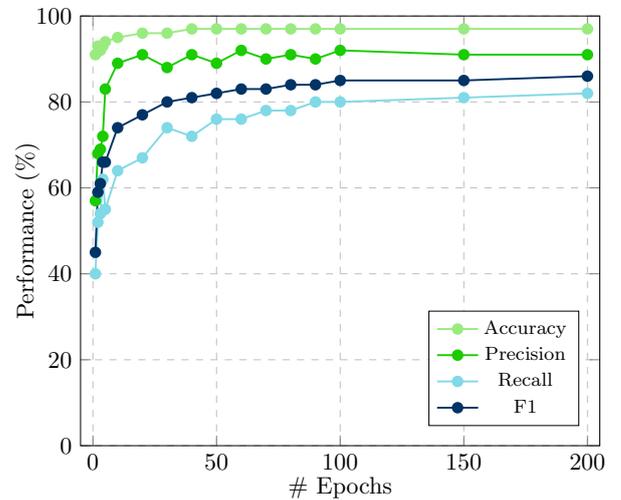
\begin{figure*}[ht!]
\begin{subfigure}{0.5\textwidth}
	\begin{center}
\begin{tikzpicture}[scale=\plotscale]
\begin{axis}[
  xlabel=\# Neurons,
  ylabel=Performance (\%),
  xmajorgrids=true,  
  ymajorgrids=true,
  grid style=dashed,
  xmin=-5, xmax=255,
  ymin=0, ymax=100,
  x label style={at={(axis description cs:0.5,0.03)}},
  y label style={at={(axis description cs:0.06,0.5)}},
  width = \columnwidth,
  legend style={font=\footnotesize,at={(0.67,0.18)},anchor=west},
   ]
\addplot[color=lightgreen,mark=*, thick] table [y=accuracy, x=neurons, col sep=comma]{plots/06_neurons.dat};
\addlegendentry{Accuracy}
\addplot[color=green,mark=*, thick] table [y=precision, x=neurons, col sep=comma]{plots/06_neurons.dat};
\addlegendentry{Precision}
\addplot[color=lightblue,mark=*, thick] table [y=recall, x=neurons, col sep=comma]{plots/06_neurons.dat};
\addlegendentry{Recall}
\addplot[color=darkblue,mark=*, thick] table [y=f1, x=neurons, col sep=comma]{plots/06_neurons.dat};
\addlegendentry{F1}
\end{axis}
\end{tikzpicture}
\caption{Results for different number of neurons.}
\label{fig:neurons}
\end{center}
\end{subfigure}
\begin{subfigure}{0.5\textwidth}
	\begin{center}
\begin{tikzpicture}[scale=\plotscale]
\begin{axis}[
  xlabel=Batch size,
  ylabel=F1 Score  (\%),
  xmajorgrids=true,  
  ymajorgrids=true,
  grid style=dashed,
  xmin=-100, xmax=5100,
  ymin=57, ymax=73,
  x label style={at={(axis description cs:0.5,0.03)}},
  y label style={at={(axis description cs:0.06,0.5)}},
  width = \columnwidth,
  legend style={font=\footnotesize,at={(0.3,0.25)},anchor=west},
   ]
\addplot[color=darkblue,mark=*, thick] table [y=f1, x=batchSize, col sep=comma]{plots/07_batch_size.dat};
\end{axis}
\end{tikzpicture}
\caption{Results for different batch sizes.}
\label{fig:batch-size}
\end{center}
\end{subfigure}
\begin{subfigure}{0.5\textwidth}
	\begin{center}
\vspace{5mm}
\begin{tikzpicture}[scale=\plotscale]
\begin{axis}[
  xlabel=Dropout Rate (\%),
  ylabel=F1 Score (\%),
  xmajorgrids=true,  
  ymajorgrids=true,
  grid style=dashed,
  xmin=-1, xmax=51,
  ymin=55, ymax=73,
  x label style={at={(axis description cs:0.5,0.03)}},
  y label style={at={(axis description cs:0.06,0.5)}},
  width = \columnwidth,
  legend style={font=\footnotesize,at={(0.3,0.25)},anchor=west},
   ]
\addplot[color=darkblue,mark=*, thick] table [y=f1, x=dropout, col sep=comma]{plots/08_dropout.dat};
\end{axis}
\end{tikzpicture}
\caption{Results for different dropout rates.}
\label{fig:dropout}
\end{center}
\end{subfigure}
\begin{subfigure}{0.5\textwidth}
	\begin{center}
\vspace{5mm}
\begin{tikzpicture}[scale=\plotscale]
\begin{axis}[
  xlabel=\# Epochs,
  ylabel=Performance (\%),
  xmajorgrids=true,  
  ymajorgrids=true,
  grid style=dashed,
  xmin=-5, xmax=205,
  ymin=0, ymax=100,
  x label style={at={(axis description cs:0.5,0.03)}},
  y label style={at={(axis description cs:0.06,0.5)}},
  width = \columnwidth,
  legend style={font=\footnotesize,at={(0.67,0.18)},anchor=west},
   ]
\addplot[color=lightgreen,mark=*, thick] table [y=accuracy, x=epochs, col sep=comma]{plots/09_epochs.dat};
\addlegendentry{Accuracy}
\addplot[color=green,mark=*, thick] table [y=precision, x=epochs, col sep=comma]{plots/09_epochs.dat};
\addlegendentry{Precision}
\addplot[color=lightblue,mark=*, thick] table [y=recall, x=epochs, col sep=comma]{plots/09_epochs.dat};
\addlegendentry{Recall}
\addplot[color=darkblue,mark=*, thick] table [y=f1, x=epochs, col sep=comma]{plots/09_epochs.dat};
\addlegendentry{F1}
\end{axis}
\end{tikzpicture}
\caption{Results for different training epochs.}
\label{fig:epochs}
\end{center}
\end{subfigure}
\caption{Influence of the hyperparameters of the word2vec embedding on the overall results (RQ-2).}
\end{figure*}

To answer RQ-3, it is necessary to determine suitable hyperparameters for the LSTM model.
When evaluating different settings of one hyperparameter of the LSTM, the baseline model is still used for all other hyperparameters (see Section~\ref{baseline}): n=5, m=200, 30 neurons, 10 epochs, dropout 20\%, and adam optimizer.
The word2vec model with the ideal configuration determined before is used for embedding the code samples.

\textbf{Number of neurons}:
A higher number of neurons allows the model to capture more complex structures, but also increases the training time.
As our results show (cf.~Figure~\ref{fig:neurons}), more neurons lead to a better performing model, with diminishing returns after around 50-70 neurons.
The training time roughly doubles going from 1 neuron to 100 neurons, and doubles again from 100 to 250 neurons.
For more epochs we reached the technical limits of our machines.
Thus, 100 neurons are chosen as the best configuration.

\textbf{Batch size}:
We tried typical batch sizes (32, 64 and 128) as well as some much smaller and much larger ones.
The results in Figure \ref{fig:batch-size} do not show that the batch size has a very strong influence on the overall performance of the model.
Only a considerable huge batch size of more than 1,000 reduces the performance.
On the contrary, the batch size influenced the time needed for training the model significantly.
While training with a batch size of 5,000 took 45s per epoch, a batch size of 200 took 130s, for a batch size of 64 it was 270s, and for a batch size of 32 around 370s, which was the smallest batch size that was feasible with respect to computation time.
At a batch size of 10, it took almost twenty minutes to train the model for just one epoch, so the training was aborted.
It can be concluded that for batch sizes smaller than 64, there is no improvement in recall that would justify putting in the extra time needed for training with such small chunks of samples.
Therefore, a batch size of 128 was considered as optimal.

\textbf{Dropout}:
Dropout and recurrent dropout are chosen together.
The baseline model is trained again, but this time for 30 epochs.
There are still some variations in the results that can account for a variance of around 2 percentage points.
As the obtained results (cf.~Figure~\ref{fig:dropout}) show, the model performs well until a dropout rate of 25\%.
However, a larger loss of neurons causes the overall performance to slowly decrease.
Therefore, it seems like a justifiable choice to set the default dropout rate to 20\%, preventing overfitting while still allowing for sufficient model performance.

\textbf{Number of training epochs}:
Training the model for more epochs increases the performance, at least up to a certain point.
The model was trained with 100 neurons.
Note that the accuracy, precision, recall and F1 score are calculated from the performance on the validation set.
Also, the model includes a dropout rate of 20\%, which should work to prevent overfitting.
Obviously, more epochs also mean that it takes longer to train the whole model.
Our results (cf.~Figure~\ref{fig:epochs}) show that there are significant improvements from longer training times.
However, there is not much to be gained after 100 epochs, and 100 epochs are chosen for the model.

\begin{table}[h!]
\begin{small}
\caption{Results for different optimizers.}
\label{tab:optimizer}
\centering
\begin{tabu}{|l|c||c|c|c|c|}	
\hline
\textbf{Optimizer} & \textbf{\# Epochs} & \textbf{Acc.} & \textbf{Prec.} & \textbf{Recall} & \textbf{F1} \\
\hline
Adam & 10 & 95\% & 85\% & 63\% & 72\% \\
Adagrad & 10 & 94\% & 78\% & 56\% & 65\% \\
Adamax & 10 & 94\% & 78\% & 56\% & 65\% \\
Nadam & 10 & 95\% & 86\% & 61\% & 71\% \\
RMSProp & 10 & 95\% & 86\% & 63\% & 73\% \\
SGD & 10 & 15\% & 10\% & 97\% & 19\% \\
\hline
Adam & 30 & 96\%& 90\%& 70\% & 79\% \\
RMSProp & 30 & 96\% & 91\% & 70\% & 80\% \\
Nadam & 30 & 96\% & 90\% & 67\% & 77\%\\
\hline
\end{tabu}
\end{small}
\end{table}

\textbf{Optimizer}:
The Keras model offers the standard Adam optimizer and some related optimizers, such as RMSprop and Adagrad as well as Nadam and Adamax.
They are all tried out to evaluate their performance.
As described in Section~\ref{optimizer}, the stochastic gradient descent (SGD) optimizer is unlikely to yield good results, as the loss function F1 is not necessarily convex.
However, out of curiosity, it is compared to the Adam family of optimizers.
Based on our results shown in Table \ref{tab:optimizer}, it seems that Adam, Nadam, and RMSprop perform slightly better than Adagrad and Adamax, possibly because they are well suited for online problems.
The performance of the SGD is even worse than expected.
The three best optimizers are compared again, this time with 50 epochs, which takes around three hours to train per optimizer.

As this is a very close call, Adam is chosen as the preferred standard optimizer.
This optimizer is most likely to be used in other research, and therefore allows for easier comparison.

\noindent
\hspace{2pt}
\fbox{\begin{minipage}{0.93\columnwidth}
\textbf{Answer for RQ-3}:
The hyperparameter settings deemed optimal given the data set and the restrictions in computational power and disk space are:
\begin{itemize}\itemsep0em
\item 100 neurons
\item training for 100 epochs
\item dropout and recurrent dropout of 20\%
\item batch size 128
\item optimizing with the adam optimizer
\end{itemize}
\end{minipage}}

\subsection{RQ-4 VUDENC effectiveness}
Using the determined ideal hyperparameters from the previous section, we trained an LSTM model for each vulnerability type in our data set, while the optimizers are set to maximize the F1 scores.
Finally, the models' performance is evaluated. This time, however, on the final test data set that the models have \textit{not seen} before.
Table \ref{tab:vulnerability-results} shows the results for each vulnerability type with the information about the sample sizes and the ratio of vulnerable code snippets found in our data set. To achieve fine granularity, each code file was divided in many small snippets, and each of those snippets is one observation.\\

\begin{table*}[h!]
\begin{small}
\caption{Results for each vulnerability type.}
\label{tab:vulnerability-results}
\centering
\begin{tabu}{|l||r|r|r||c|c|c|c|}
\hline 	
\multirow{2}[2]{*}{\textbf{Vulnerability}} &
\multicolumn{2}{c|}{\textbf{\# Snippets}}  &
\multirow{2}[2]{*}{\centering\textbf{Vulnerable Code Snippets}} &
\multirow{2}[2]{*}{\centering\textbf{Accuracy}} &
\multirow{2}[2]{*}{\centering\textbf{Precision}} &
\multirow{2}[2]{*}{\centering\textbf{Recall}} &
\multirow{2}[2]{*}{\centering\textbf{F1}} \\
& \textbf{Training} & \textbf{Test} &&&&& \\
\hline
SQL injection & 42,690 & 8196 & 19.2 \% & 92.5 \% & 82.2 \% & 78.0 \% & 80.1 \% \\
XSS & 8277 & 717 & 8.7 \% & 97.8 \% & 91.9 \% & 80.8 \% & 86.0 \% \\
Command injection & 18,814 & 2287 & 12.2 \% & 97.8 \%  & 94.0 \% & 87.2 \% & 90.5\%  \\
XSRF & 27434 & 3,600 & 13.1 \% & 97.2 \% & 92.9 \% & 85.4 \% & 89.0 \% \\
Remote code execution & 14412 & 1303 & 9.0 \% & 98.1 \% & 96.0 \% & 82.2 \% & 88.8 \%\\
Path disclosure & 19,680 & 2315 & 11.8 \% & 97.3 \% & 92.0 \% & 84.4 \% & 88.1 \% \\
Open redirect & 12740 & 1691 & 13.3 \% & 96.8 \% & 91.0 \% & 83.9 \% & 87.3 \%  \\
\hline
\end{tabu}
\end{small}
\end{table*}

\noindent
\hspace{2pt}
\fbox{\begin{minipage}{0.93\columnwidth}
\textbf{Answer for RQ-4}:
The precision (the fraction of true positives in all positive predictions) ranged from 82\% to 96\%, suggesting a very low false positive rate.
The recall is a little lower, between 78\% and 87\%, which means that just 13-22\% of the samples labeled as \textit{vulnerable} were missed.
Finally, the overall F1 score, the harmonic mean of precision and recall, ranges from 80\%-90\%, which is a very satisfying result.
\end{minipage}}

\section{Discussion}
\label{sec:discussion}
In this section, we interpret \VUDENC's empirical results, and discuss limitations and threats to validity of our experiments.

\subsection{Interpretation of the empirical results}
	
To provide a reference frame for the evaluation of this work, Table~\ref{tab:result comparison} summarizes the comparison of \VUDENC with its most closely related approaches. Since there are fundamental differences among the approaches, no direct comparison can be drawn between experimental results.
	Thus, the comparison covers the following aspects:
	\begin{itemize}
		\item \textbf{Language:} What language is subject of the classification efforts.
		\item \textbf{Data basis:} Does the data stem from real-life projects or from synthetic databases (such as benchmark datasets).
		\item \textbf{Labels:} How are the labels for the training data originally generated. 
		\item \textbf{Granularity:} Is the code evaluated on a rough granularity (whole classes or files) or a fine granularity (lines or tokens).
		\item \textbf{Machine Learning Approach:} What class of neural network or machine learning approach is used (CNN, RNN, LSTM).
		\item \textbf{Vulnerability types:} Which kinds of vulnerabilities are detected.
		\item \textbf{Size of dataset:} How many functions, projects, classes etc. make up the dataset.
		\item \textbf{Scope and applicability:} Has the model been trained on a single project and can it only classify files within that project, or is it generally applicable to any code from a large variety of sources.
	\end{itemize}
	
Russel et al.~\cite{Russell.2018} work on lexed source code and use a similar approach to \VUDENC. They take C/C++ code from real software packages and from benchmark datasets to find vulnerabilities. In contrast to \VUDENC, they do not use LSTMs, but the very similar CNN and RNN networks. The main difference is that they use static parsers to generate labels, while \VUDENC relies only on commit contexts. They identify five different types of vulnerabilities, including buffer overflows and null pointers, which are quite different from the vulnerabilities found in Python code. They achieve an F1 score of 0.566.
	
Pang et al.~\cite{Pang.2015} use a hybrid n-gram and feature selection analysis on a dataset consisting of four Java Android applications. Labels were obtained from a pre-prepared online benchmark. They conduct their analysis on the level of whole classes, and prediction is performed by training a support vector machine. While they achieve good results for predictions within the same project (accuracy, precision, and recall around 90\%), the cross-project results are less strong with an F1 score of around 65\%. The two main differences here are that \VUDENC uses a much larger and more diverse dataset, and that it can point out specific locations in the code that might be vulnerable.
	
The tool VuRLE~\cite{Ma.2017} was trained on 40 Java applications collected from GitHub. The commits are analyzed manually to find vulnerabilities of five different types, including SQL injections. ASTs are the basis for the approach in which ``edit groups'' are created to classify vulnerabilities using 10-fold cross-validation instead of a neural network. Although their main goal is to create repair templates, they first have to detect vulnerabilities, which they manage with an F1 score of around 65\%. However, as opposed to \VUDENC's fully automated approach, a manual classification as performed by Ma et al.\ is not feasible for larger datasets.

\begin{table*}[ht!]
\caption{Comparison of the results with those of the most similar approaches.}
    \label{tab:result comparison}
    \centering
  \begin{small}
	\begin{tabular}{ | p{3.0cm} | p{1.2cm}|  p{0.7cm}| p{1.4cm} |  p{1.3cm} | p{1.9cm} | p{1.1cm} || p{0.6cm}|p{0.6cm}|p{0.6cm}|p{0.4cm}|  }
		\hline
		\multicolumn{7}{|c||}{{\bf Characteristics of approach}} & \multicolumn{4}{c|}{{\bf Resulting metrics}} \\
		\hline
		Name &  Language & Data & Labels & Scope &Granularity & Method & Acc. & Pre. & Rec. & F1  \\
		\hline
		Russel et al.~\cite{Russell.2018} & C/C++ & real \& synth. & static analysis tool & general & token level (fine) & CNN, RNN &  &   &   &  57\%  \\
		\hline
		Pang et al.~\cite{Pang.2015} & Java & real  & pre-existing  & 4 apps & whole classes & SVM & 63\% & 67\%  & 63\%  & 65\%    \\
		\hline
		VuRLE~\cite{Ma.2017}  & Java & real  & manually identified  & general & edits (fine) & 10-fold CV &  & 65\%  & 66\%  & 65\%    \\
		\hline
		VulDeePecker~\cite{Li.2018} & C/C++ & real \& synth.  & patches \& manual & general & API/function calls & BLSTM &  &   &  & 85\%-95\%    \\
		\hline
		Dam et al.~\cite{Dam.2017}& Java & real & static analysis tool & 18 apps & whole file & LSTM & \multicolumn{4}{c|}{ 4 / 17 projects (see above)}   \\
		\hline
		Hovsepyan et al.~\cite{Hovsepyan.2012} & Java & real  & static analysis tool  & 1 project & whole file & grid search & 87\% & 	85\%  & 88\%  & 85\%   \\
		\hline
		\VUDENC & Python & real  &patches& general  & token level & LSTM & 92-98\% & 82-96\% & 78-87\% & 80-90\%   \\
		\hline
		\hline
	\end{tabular}\\
\end{small}
\end{table*}

Li et al.~\cite{Li.2018} developed the tool VulDeePecker to detect buffer error and resource management error vulnerabilities in C/C++ programs. They work on a ``code gadget database'' made from a large number of popular open-source projects, including the Linux kernel and Firefox. The vulnerabilities are found by using the NVD and SARD dataset, which contain synthetic and real-life code. Like \VUDENC, they use patches to create labels, but include a second step of manually double-checking every positive label. They train a bidirectional LSTM and achieve an F1 score of around 85-95\%, which is slightly better than our results. However, \VUDENC uses data from all kinds of real-life GitHub projects, which are not as perfectly categorized as the high-profile projects which are used to train VulDeePecker. 

Dam et al.~\cite{Dam.2017} train an LSTM on 18 Android applications. Similar to \VUDENC, they look at code tokens and represent them by multi-dimensional vectors. However, their feature extraction algorithm is focused on the specific project, learning relevant syntactic and semantic information that is specific for this single application. In their best configuration, they achieve a F1 measure of around 91\%---however, only for predictions within the \textit{same} project on which the classifier was trained. For inter-project predictions, they do not report those metrics, but rather state that they reached precision and recall of more than 80\% for only 4 out of 17 projects. Moreover, they only classify whole files, working on a very rough granularity.
	
Hovsepyan et al.~\cite{Hovsepyan.2012} focus their efforts on one single Java application and try to predict whether a whole file is vulnerable or not. To this end, they split the file in tokens and use a static analysis tool to create the labels for their files, and use a radial base function and a grid search algorithm to learn their features for the classification. They achieve an accuracy, precision, and recall of over 80\%. It is unclear how useful those results are, given that they work only on this single project, and they cannot locate vulnerabilities, but only classify whole files.
	
In conclusion, higher precision, recall and F1 scores are easier to achieve when focusing on predictions within a single project and on the granularity level of files. Training a classifier that is applicable for general detection of vulnerabilities and for pointing out their exact location is much harder, but also tend to lead to a more useful result. 
Furthermore, the quality of a model heavily depends on the quality of the underlying data. Classification and prediction on synthetic datasets, or datasets that are curated and selected manually, is easier than training a model on real-life code from sources such as GitHub. With essentially the same approach, Russel et al.~\cite{Russell.2018} achieved 57\% on natural code from GitHub, but 84\% on the SATE test suite due to its clean and consistent style and structure. The \VUDENC approach arguably performs quite well, given that it works purely on natural real-life code and serves as a general vulnerability detector that can be used at the granularity level of code tokens.
However, as already mentioned, a direct comparison with related approaches is not possible and was not even the aim of our experiments. The most important evaluation goal is to show that it is possible to make use of a large dataset of real-life source code in order to fully automatically train a vulnerability prediction model that can be applied across project boundaries, while achieving an accuracy which is comparable or even better than those of related approaches.

\subsection{Limitations and threats to validity}
	
{\bf Labeling based on the commit context:}
	Our approach relies on commit contexts for classifying code snippets into vulnerable or (probably) not vulnerable, assuming that there was an actual vulnerability and the fixed version is in some way better than the previous one. 
	However, sanity checks showed that there are cases in which a fix does not solve a problem, there are several vulnerabilities at the same time, or even a new vulnerability is introduced. 
	In essence, there may be errors in the training data presented to \VUDENC, and there is no automated method of double-checking. However, we work on substantially large datasets in which those issues mostly fade to the background because of the large number of legitimate fixes.
	Moreover, our main focus is easy automation without the need for human expert oversight, and to work out the insights that can be gained without \textit{any} prior knowledge about vulnerabilities, just using the code database as a source of information for the neural network.  
	
{\bf Oversights due to developer decisions:}
	Our training data is limited to fixes that were actually applied by developers. Undetected vulnerabilities or vulnerabilities being ignored by developers remain invisible, leading to blind spots in the trained classifier. Liu et al.~\cite{Liu.2018} exploit this \textit{on purpose}, as they are trying to learn which violations are actually fixed by developers and which thus are, most likely, true positives. The situation is therefore ambivalent: While false positives can be avoided by only taking problems into account that were actually fixed, false negatives could be introduced for vulnerabilities that developers do not notice, understand or care about.

{\bf Unknown vulnerabilities:}
	A prerequisite for our approach is the existence of known vulnerabilities to learn from. In cases where no known examples for code with vulnerabilities are available, our method cannot be applied. 
	In the context of vulnerability detection, however, as pointed out by Yamaguchi et al.~\cite{Yamaguchi.2012}, such situations are rare in practice, as the main concern for large software repositories is not to discover a single vulnerability, but to make sure that the same type of error does not spread across projects, which is what our method is useful for. 

{\bf Capturing the context of a vulnerability:}
	The diff describing the change provides the changed lines and three lines before and after them, respectively. However, there may be situations in which a vulnerability stems from the interaction of lines of code that are spread over a large file or several files. Our model cannot learn the implications of such far-reaching dependencies.

{\bf Types of vulnerabilities:}
	Although we took a variety of different types of vulnerabilities into account, even more than in many other works, the list is not meant to be complete. Our work focuses on the most important and typical ones for which we are confident to train our model on a substantial dataset.
	
{\bf Usage of open-source data:}
	The data collected from GitHub might not be entirely representative. In particular, our findings might not be applicable to closed-source projects applying different quality assurance methods. Furthermore, although projects that were duplicates of each other have been excluded, it is possible that several projects in the database are very similar to each other. By the nature of open-source and code sharing practices, some parts of the code in different repositories might even come close to being duplicates, meaning that the dataset is more limited than it first seems to be. 

{\bf Dealing with Python source code:}
	The present design of this work is limited to dealing with vulnerabilities in Python source code. The detection of vulnerabilities in binary files or executables is a different problem that is not tackled here. Moreover, only Python files were taken into account, while others such as configuration files written in XML were ignored, although they might include security-relevant information. In principle, however, the approach taken by \VUDENC is not restricted to Python, but could be applied to other programming or markup languages.

{\bf Evaluation of performance predictions:}
To evaluate the performance of the predictions, we use the F1 score 
which provides a balanced score that takes precision and recall into 
account. In practice, however, there may be other measures demonstrating how well a classifier performs, which threatens the conclusion validity of our results. Nonetheless, the chosen measures are standard performance measures that have been applied in other works regarding vulnerability prediction, thus threats to conclusion validity are minimized.
Finally, because of the non-deterministic nature of the training process, the same model can be trained on the same data two times, one directly after the other, and the resulting scores for precision, recall etc. can diverge by a few percentage points. However, since training one specific configuration already took one to ten hours to be completed, we abstained from several repetitions of the training process.

\section{Conclusion}
\label{sec:conclusion}
In this article, we presented \VUDENC, a system for vulnerability detection based on deep learning on a natural codebase. \VUDENC's purpose is to relieve human experts from the time-consuming and subjective work of manually defining features for vulnerability detection. This work demonstrates the potential of using machine learning directly on source code to learn such vulnerability features by leveraging LSTM models.

To create the basis for \VUDENC, a large dataset of commits was mined from GitHub and labeled according to the commit context. The data stems from several hundred real-world repositories containing natural Python source code and covers seven different types of vulnerabilities, including SQL injections, cross-site scripting and command injections. The dataset has been made publicly available and can be used to replicate this work and conduct further research. A word2vec model has been trained on a large corpus of Python code to be able to perform embeddings of code tokens that preserve semantics, and has been made available as well.
	
The raw source code was preprocessed and the datasets for each vulnerability were built by taking every single code token with its context (the tokens before and after it) as one sample and embedding it using the word2vec model. The LSTM network was trained to detect vulnerable code on the level of individual tokens.
	
Systematic experiments show that \VUDENC achieves a recall of 78\%-87\%, a precision of 82-96\% and an F1 score of 80-90\%.  These results are very promising and encourage further research in this area. \VUDENC is able to highlight the specific areas in code that are likely to contain vulnerabilities and provide confidence levels for its predictions. It can be adjusted to focus on minimizing the rate of false positives or false negatives.
	
Future work should focus on improving understandability and actionability of \VUDENC's results as these are the main characteristics \cite{Morrison.2015,TanTDM15} for the practical applicability of a bug or vulnerability prediction tools. Furthermore, the overall approach could be improved with respect to labeling the data, combining \VUDENC with other approaches for enhanced results, and leveraging the commit context to create actionable fix recommendations. Additionally, \VUDENC could also be extended to other programming languages or types of vulnerabilities, and the word2vec model could be replaced with programming language specific code2vec models~\cite{AlonZLY19,WangS2020} that also also consider detailed AST-like features.
	
The code for \VUDENC has been made available as a public GitHub repository~\cite{Wartschinski.2.12.2019b} alongside with trained models, and examples. All data, including the training sets for the python word2vec model and the actual datasets for the vulnerabilities, can be found on the Zenodo platform \cite{Wartschinski.2.12.2019, Wartschinski.1.12.2019, Wartschinski.2.12.2019c}.

\bibliography{VulnerabilityDetection}

\begin{thebibliography}{10}
\expandafter\ifx\csname url\endcsname\relax
  \def\url#1{\texttt{#1}}\fi
\expandafter\ifx\csname urlprefix\endcsname\relax\def\urlprefix{URL }\fi
\expandafter\ifx\csname href\endcsname\relax
  \def\href#1#2{#2} \def\path#1{#1}\fi

\bibitem{DanGoodin.2017}
{Dan Goodin},
  \href{https://arstechnica.com/information-technology/2017/05/an-nsa-derived-ransomware-worm-is-shutting-down-computers-worldwide/}{An
  {NSA}-derived ransomware worm is shutting down computers worldwide} (2017)
  [cited 21.12.2020].
\newline\urlprefix\url{https://arstechnica.com/information-technology/2017/05/an-nsa-derived-ransomware-worm-is-shutting-down-computers-worldwide/}

\bibitem{Yamaguchi.2012}
F.~Yamaguchi, M.~Lottmann, K.~Rieck, Generalized vulnerability extrapolation
  using abstract syntax trees, in: Proceedings of the 28th Annual Computer
  Security Applications Conference, 2012, pp. 359--368.

\bibitem{Durumeric.2014}
Z.~Durumeric, F.~Li, J.~Kasten, J.~Amann, J.~Beekman, M.~Payer, N.~Weaver,
  D.~Adrian, V.~Paxson, M.~Bailey, et~al., The matter of heartbleed, in:
  Proceedings of the 2014 conference on internet measurement conference, 2014,
  pp. 475--488.

\bibitem{jurjens2002umlsec}
J.~J{\"u}rjens, Umlsec: Extending uml for secure systems development, in:
  International Conference on The Unified Modeling Language, Springer, 2002,
  pp. 412--425.

\bibitem{peldszus2021ontology}
S.~Peldszus, J.~B{\"u}rger, T.~Kehrer, J.~J{\"u}rjens, Ontology-driven
  evolution of software security, Data \& Knowledge Engineering (2021) 101907.

\bibitem{burger2020ontology}
J.~B{\"u}rger, T.~Kehrer, J.~J{\"u}rjens, Ontology evolution in the context of
  model-based secure software engineering, in: International Conference on
  Research Challenges in Information Science, Springer, 2020, pp. 437--454.

\bibitem{CVE}
CVE, \href{http://cve.mitre.org/}{Cve} (2019) [cited 21.12.2020].
\newline\urlprefix\url{http://cve.mitre.org/}

\bibitem{arkin2005software}
B.~Arkin, S.~Stender, G.~McGraw, Software penetration testing, IEEE Security \&
  Privacy 3~(1) (2005) 84--87.

\bibitem{Zimmermann.2010}
T.~Zimmermann, N.~Nagappan, L.~Williams, Searching for a needle in a haystack:
  Predicting security vulnerabilities for windows vista, in: 2010 Third
  International Conference on Software Testing, Verification and Validation,
  2010, pp. 421--428.

\bibitem{Shin.2010}
Y.~Shin, A.~Meneely, L.~Williams, J.~A. Osborne, Evaluating complexity, code
  churn, and developer activity metrics as indicators of software
  vulnerabilities, IEEE Transactions on Software Engineering 37~(6) (2010)
  772--787.

\bibitem{Shin.2013}
Y.~Shin, L.~Williams, Can traditional fault prediction models be used for
  vulnerability prediction?, Empirical Software Engineering 18~(1) (2013)
  25--59.

\bibitem{kim2017vuddy}
S.~Kim, S.~Woo, H.~Lee, H.~Oh, Vuddy: A scalable approach for vulnerable code
  clone discovery, in: 2017 IEEE Symposium on Security and Privacy (SP), IEEE,
  2017, pp. 595--614.

\bibitem{li2016vulpecker}
Z.~Li, D.~Zou, S.~Xu, H.~Jin, H.~Qi, J.~Hu, Vulpecker: an automated
  vulnerability detection system based on code similarity analysis, in:
  Proceedings of the 32nd Annual Conference on Computer Security Applications,
  2016, pp. 201--213.

\bibitem{Scandariato.2014}
R.~Scandariato, J.~Walden, A.~Hovsepyan, W.~Joosen, Predicting vulnerable
  software components via text mining, IEEE Transactions on Software
  Engineering 40~(10) (2014) 993--1006.

\bibitem{Morrison.2015}
P.~Morrison, K.~Herzig, B.~Murphy, L.~Williams, Challenges with applying
  vulnerability prediction models, in: Proceedings of the 2015 Symposium and
  Bootcamp on the Science of Security, 2015, p.~4.

\bibitem{Russell.2018}
R.~Russell, L.~Kim, L.~Hamilton, T.~Lazovich, J.~Harer, O.~Ozdemir,
  P.~Ellingwood, M.~McConley, Automated vulnerability detection in source code
  using deep representation learning, in: 2018 17th IEEE International
  Conference on Machine Learning and Applications (ICMLA), 2018, pp. 757--762.

\bibitem{Pang.2015}
Y.~Pang, X.~Xue, A.~S. Namin, Predicting vulnerable software components through
  n-gram analysis and statistical feature selection, in: 2015 IEEE 14th
  International Conference on Machine Learning and Applications (ICMLA), 2015,
  pp. 543--548.

\bibitem{Ma.2017}
S.~Ma, F.~Thung, D.~Lo, C.~Sun, R.~H. Deng, Vurle: Automatic vulnerability
  detection and repair by learning from examples, in: European Symposium on
  Research in Computer Security, 2017, pp. 229--246.

\bibitem{Li.2018}
Z.~Li, D.~Zou, S.~Xu, X.~Ou, H.~Jin, S.~Wang, Z.~Deng, Y.~Zhong, Vuldeepecker:
  {A} deep learning-based system for vulnerability detection, in: 25th Annual
  Network and Distributed System Security Symposium, {NDSS} 2018, San Diego,
  California, USA, February 18-21, 2018, The Internet Society, 2018.

\bibitem{Dam.2017}
H.~K. Dam, T.~Tran, T.~Pham, S.~W. Ng, J.~Grundy, A.~Ghose, Automatic feature
  learning for vulnerability prediction, arXiv preprint arXiv:1708.02368.

\bibitem{Hovsepyan.2012}
A.~Hovsepyan, R.~Scandariato, W.~Joosen, J.~Walden, Software vulnerability
  prediction using text analysis techniques, in: Proceedings of the 4th
  international workshop on Security measurements and metrics, 2012, pp. 7--10.

\bibitem{austin2011one}
A.~Austin, L.~Williams, One technique is not enough: A comparison of
  vulnerability discovery techniques, in: 2011 International Symposium on
  Empirical Software Engineering and Measurement, IEEE, 2011, pp. 97--106.

\bibitem{ceccato2016static}
M.~Ceccato, R.~Scandariato, Static analysis and penetration testing from the
  perspective of maintenance teams, in: Proceedings of the 10th ACM/IEEE
  International Symposium on Empirical Software Engineering and Measurement,
  2016, pp. 1--6.

\bibitem{Ghaffarian.2017}
S.~M. Ghaffarian, H.~R. Shahriari, Software vulnerability analysis and
  discovery using machine-learning and data-mining techniques: A survey, ACM
  Computing Surveys (CSUR) 50~(4) (2017) 56.

\bibitem{Hochreiter.1997}
S.~Hochreiter, J.~Schmidhuber, Long short-term memory, Neural computation 9~(8)
  (1997) 1735--1780.

\bibitem{Word2Vec}
{Jay Alammar}, \href{https://jalammar.github.io/illustrated-word2vec/}{The
  illustrated word2vec} (27. 03. 2019) [cited 21.12.2020].
\newline\urlprefix\url{https://jalammar.github.io/illustrated-word2vec/}

\bibitem{Liu.2018}
K.~Liu, D.~Kim, T.~F. Bissyand{\'e}, S.~Yoo, Y.~{Le Traon}, Mining fix patterns
  for findbugs violations, IEEE Transactions on Software Engineering.

\bibitem{Hall.2011}
T.~Hall, S.~Beecham, D.~Bowes, D.~Gray, S.~Counsell, A systematic literature
  review on fault prediction performance in software engineering, IEEE
  Transactions on Software Engineering 38~(6) (2011) 1276--1304.

\bibitem{Nagappan+2006}
N.~Nagappan, T.~Ball, A.~Zeller,
  \href{https://doi.org/10.1145/1134285.1134349}{Mining metrics to predict
  component failures}, in: Proceedings of the 28th International Conference on
  Software Engineering, ICSE ’06, ACM, 2006, pp. 452--461.
\newblock \href {http://dx.doi.org/10.1145/1134285.1134349}
  {\path{doi:10.1145/1134285.1134349}}.
\newline\urlprefix\url{https://doi.org/10.1145/1134285.1134349}

\bibitem{Nagappan.2008}
N.~Nagappan, B.~Murphy, V.~Basili, The influence of organizational structure on
  software quality, in: 2008 ACM/IEEE 30th International Conference on Software
  Engineering, 2008, pp. 521--530.

\bibitem{Shin.2008}
Y.~Shin, L.~Williams, An empirical model to predict security vulnerabilities
  using code complexity metrics, in: Proceedings of the Second ACM-IEEE
  international symposium on Empirical software engineering and measurement,
  2008, pp. 315--317.

\bibitem{Chowdhury.2011}
I.~Chowdhury, M.~Zulkernine, Using complexity, coupling, and cohesion metrics
  as early indicators of vulnerabilities, Journal of Systems Architecture
  57~(3) (2011) 294--313.

\bibitem{Neuhaus.2007}
S.~Neuhaus, T.~Zimmermann, C.~Holler, A.~Zeller, Predicting vulnerable software
  components, in: ACM Conference on computer and communications security, 2007,
  pp. 529--540.

\bibitem{Zhou.2017}
Y.~Zhou, A.~Sharma, Automated identification of security issues from commit
  messages and bug reports, in: Proceedings of the 2017 11th Joint Meeting on
  Foundations of Software Engineering, 2017, pp. 914--919.

\bibitem{Li.2005}
Z.~Li, Y.~Zhou, Pr-miner: automatically extracting implicit programming rules
  and detecting violations in large software code, in: ACM SIGSOFT Software
  Engineering Notes, Vol.~30, 2005, pp. 306--315.

\bibitem{Grieco.2016}
G.~Grieco, G.~L. Grinblat, L.~Uzal, S.~Rawat, J.~Feist, L.~Mounier, Toward
  large-scale vulnerability discovery using machine learning, in: Proceedings
  of the Sixth ACM Conference on Data and Application Security and Privacy,
  2016, pp. 85--96.

\bibitem{Yamaguchi.2011}
F.~Yamaguchi, F.~Lindner, K.~Rieck, Vulnerability extrapolation: assisted
  discovery of vulnerabilities using machine learning, in: Proceedings of the
  5th USENIX conference on Offensive technologies, 2011, p.~13.

\bibitem{Shar.2013b}
L.~K. Shar, H.~B.~K. Tan, Predicting sql injection and cross site scripting
  vulnerabilities through mining input sanitization patterns, Information and
  Software Technology 55~(10) (2013) 1767--1780.

\bibitem{Shar.2013}
L.~K. Shar, H.~B.~K. Tan, L.~C. Briand, Mining sql injection and cross site
  scripting vulnerabilities using hybrid program analysis, in: Proceedings of
  the 2013 International Conference on Software Engineering, 2013, pp.
  642--651.

\bibitem{Wang.2016}
S.~Wang, T.~Liu, L.~Tan, Automatically learning semantic features for defect
  prediction, in: 2016 IEEE/ACM 38th International Conference on Software
  Engineering (ICSE), 2016, pp. 297--308.

\bibitem{Gupta.2017b}
R.~Gupta, S.~Pal, A.~Kanade, S.~Shevade, Deepfix: Fixing common c language
  errors by deep learning, in: Thirty-First AAAI Conference on Artificial
  Intelligence, 2017.

\bibitem{Dam.2016b}
H.~K. Dam, T.~Tran, J.~Grundy, A.~Ghose, Deepsoft: A vision for a deep model of
  software, in: Proceedings of the 2016 24th ACM SIGSOFT International
  Symposium on Foundations of Software Engineering, 2016, pp. 944--947.

\bibitem{Harer.2018}
J.~Harer, O.~Ozdemir, T.~Lazovich, C.~Reale, R.~Russell, L.~Kim, et~al.,
  Learning to repair software vulnerabilities with generative adversarial
  networks, in: Advances in Neural Information Processing Systems, 2018, pp.
  7933--7943.

\bibitem{Gupta.2017}
R.~Gupta, S.~Pal, A.~Kanade, S.~Shevade, Deepfix: Fixing common c language
  errors by deep learning, in: Thirty-First AAAI Conference on Artificial
  Intelligence, 2017.

\bibitem{Bellon.2007}
S.~Bellon, R.~Koschke, G.~Antoniol, J.~Krinke, E.~Merlo, Comparison and
  evaluation of clone detection tools, IEEE Transactions on Software
  Engineering 33~(9) (2007) 577--591.

\bibitem{Rolim.2018}
R.~Rolim, G.~Soares, R.~Gheyi, T.~Barik, L.~D'Antoni, Learning quick fixes from
  code repositories, arXiv preprint arXiv:1803.03806.

\bibitem{AyeshaCuthbert.15.4.2019}
{Ayesha Cuthbert},
  \href{https://hackernoon.com/8-top-programming-languages-frameworks-of-2019-2f08d2d21a1}{8
  top programming languages {\&} frameworks of 2019} (2019) [cited 21.12.2020].
\newline\urlprefix\url{https://hackernoon.com/8-top-programming-languages-frameworks-of-2019-2f08d2d21a1}

\bibitem{VidushiDwivedi.}
{Vidushi Dwivedi},
  \href{https://www.geeksforgeeks.org/top-10-programming-languages-of-the-world-2019-to-begin-with/}{Top
  10 programming languages of the world} (2019) [cited 21.12.2020].
\newline\urlprefix\url{https://www.geeksforgeeks.org/top-10-programming-languages-of-the-world-2019-to-begin-with/}

\bibitem{Github.com.19}
Github.com, \href{https://octoverse.github.com/projects}{Github - the state of
  the octoverse 2018} (2019) [cited 21.12.2020].
\newline\urlprefix\url{https://octoverse.github.com/projects}

\bibitem{OWASPFoundation.}
{OWASP Foundation} [cited 21.12.2020].
\newblock \href{https://www.owasp.org}{[link]}.
\newline\urlprefix\url{https://www.owasp.org}

\bibitem{Medeiros.2014}
I.~Medeiros, N.~F. Neves, M.~Correia, Automatic detection and correction of web
  application vulnerabilities using data mining to predict false positives, in:
  Proceedings of the 23rd international conference on World wide web, 2014, pp.
  63--74.

\bibitem{TanTDM15}
M.~Tan, L.~Tan, S.~Dara, C.~Mayeux, Online defect prediction for imbalanced
  data, in: 37th {IEEE/ACM} International Conference on Software Engineering,
  {ICSE} 2015, {IEEE} Computer Society, 2015, pp. 99--108.

\bibitem{Wartschinski.2.12.2019b}
L.~Wartschinski,
  \href{https://github.com/LauraWartschinski/VulnerabilityDetection}{Vudenc~github
  repository} (2020) [cited 21.12.2020].
\newline\urlprefix\url{https://github.com/LauraWartschinski/VulnerabilityDetection}

\bibitem{Wartschinski.2.12.2019c}
L.~Wartschinski, \href{https://zenodo.org/record/3559841#.XeVaZNVG2Hs}{Vudenc~-
  datasets for vulnerabilities} (2020) [cited 21.12.2020].
\newblock \href {http://dx.doi.org/10.5281/zenodo.3559841}
  {\path{doi:10.5281/zenodo.3559841}}.
\newline\urlprefix\url{https://zenodo.org/record/3559841#.XeVaZNVG2Hs}

\bibitem{Dam.2016}
H.~K. Dam, T.~Tran, T.~Pham, A deep language model for software code, arXiv
  preprint arXiv:1608.02715.

\bibitem{Hindle.2012}
A.~Hindle, E.~T. Barr, Z.~Su, M.~Gabel, P.~Devanbu, On the naturalness of
  software, in: 2012 34th International Conference on Software Engineering
  (ICSE), 2012, pp. 837--847.

\bibitem{Allamanis.2018}
M.~Allamanis, E.~T. Barr, P.~Devanbu, C.~Sutton, A survey of machine learning
  for big code and naturalness, ACM Computing Surveys (CSUR) 51~(4) (2018) 81.

\bibitem{Kumar.2019}
H.~Kumar, B.~Ravindran, Polyphonic music composition with lstm neural networks
  and reinforcement learning, arXiv preprint arXiv:1902.01973.

\bibitem{Tu.2014}
Z.~Tu, Z.~Su, P.~Devanbu, On the localness of software, in: Proceedings of the
  22nd ACM SIGSOFT International Symposium on Foundations of Software
  Engineering, 2014, pp. 269--280.

\bibitem{Kingma.2014}
D.~P. Kingma, J.~Ba, Adam: {A} method for stochastic optimization, in:
  Y.~Bengio, Y.~LeCun (Eds.), 3rd International Conference on Learning
  Representations, {ICLR} 2015, San Diego, CA, USA, May 7-9, 2015, Conference
  Track Proceedings, 2015.

\bibitem{Wartschinski.1.12.2019}
L.~Wartschinski, \href{https://zenodo.org/record/3559203#.XeRoytVG2Hs}{Vudenc -
  dataset with diff files} (2020) [cited 21.12.2020].
\newblock \href {http://dx.doi.org/10.5281/zenodo.3559203}
  {\path{doi:10.5281/zenodo.3559203}}.
\newline\urlprefix\url{https://zenodo.org/record/3559203#.XeRoytVG2Hs}

\bibitem{Wartschinski.2.12.2019}
L.~Wartschinski, \href{https://zenodo.org/record/3559480#.XeTMzdVG2Hs}{Vudenc~-
  python corpus for word2vec} (2020) [cited 21.12.2020].
\newblock \href {http://dx.doi.org/10.5281/zenodo.3559480}
  {\path{doi:10.5281/zenodo.3559480}}.
\newline\urlprefix\url{https://zenodo.org/record/3559480#.XeTMzdVG2Hs}

\bibitem{OWASP_SQLInjection}
\href{https://owasp.org/www-community/attacks/SQL_Injection}{{SQL Injection}}
  (2020) [cited 06.07.2021].
\newline\urlprefix\url{https://owasp.org/www-community/attacks/SQL_Injection}

\bibitem{OWASP_XXS}
\href{https://owasp.org/www-community/attacks/xss/}{{Cross-site Scripting}}
  (2020) [cited 06.07.2021].
\newline\urlprefix\url{https://owasp.org/www-community/attacks/xss/}

\bibitem{OWASP_CommandInjection}
\href{https://owasp.org/www-community/attacks/Command_Injection}{{Command
  Injection}} (2020) [cited 06.07.2021].
\newline\urlprefix\url{https://owasp.org/www-community/attacks/Command_Injection}

\bibitem{OWASP_XSRF}
\href{https://owasp.org/www-community/attacks/csrf}{{Cross Site Request
  Forgery}} (2020) [cited 06.07.2021].
\newline\urlprefix\url{https://owasp.org/www-community/attacks/csrf}

\bibitem{OWASP_CodeInjection}
\href{https://owasp.org/www-community/attacks/Code_Injection}{{Code Injection}}
  (2020) [cited 06.07.2021].
\newline\urlprefix\url{https://owasp.org/www-community/attacks/Code_Injection}

\bibitem{CWE_OpenRedirect}
\href{https://cwe.mitre.org/data/definitions/601.html}{{CWE-601: Open
  Redirect}} (2020) [cited 06.07.2021].
\newline\urlprefix\url{https://cwe.mitre.org/data/definitions/601.html}

\bibitem{AlonZLY19}
U.~Alon, M.~Zilberstein, O.~Levy, E.~Yahav, code2vec: learning distributed
  representations of code, Proc. {ACM} Program. Lang. 3~({POPL}) (2019)
  40:1--40:29.

\bibitem{WangS2020}
K.~Wang, Z.~Su, Blended, precise semantic program embeddings, in: Proceedings
  of the 41st {ACM} {SIGPLAN} International Conference on Programming Language
  Design and Implementation, {PLDI} 2020, {ACM}, 2020, pp. 121--134.

\end{thebibliography}
\end{document}